\magnification=\magstep1
\font\fourteenbf=cmbx10 scaled\magstep 2
\font\twelvebf=cmbx10 scaled\magstep 1
 1
\font\twelveit=cmti10 scaled\magstep 1
\font\elevenit=cmti8 scaled\magstep 0
\font\eightrm=cmr8

\newskip\bottommarkspace
\newskip\normparindent
\newbox\AUTHORBOX\newbox\ADDRESSBOX
\newbox\titlebox\newbox\abstractbox\newbox\footnotebox
\newbox\datebox\newbox\authorbox\newbox\brolbox\newskip\titleskip
\newbox\CHAPTERBOX\newbox\SECTIONBOX
\newdimen\broldimen
\newcount\CHAPNUMBER\CHAPNUMBER=0
\newcount\SECTNUMBER\SECTNUMBER=0
\newcount\SECTNUMBERA\SECTNUMBERA=0
\newcount\SECTNUMBERB\SECTNUMBERB=0
\newcount\SECTNUMBERC\SECTNUMBERC=0
\newcount\FORMCOUNTER\FORMCOUNTER=0
\newcount\REFCOUNTER\REFCOUNTER=0
\newcount\EQACOUNTER\EQACOUNTER=0
\newcount\ITEMCOUNTER\ITEMCOUNTER=0
\newcount\ITEMITEMCOUNTER\ITEMITEMCOUNTER=0
\newcount\APPCOUNTER\APPCOUNTER=-1
\newcount\chapnumber\chapnumber=\the\CHAPNUMBER
\newcount\refcounter\refcounter=\the\REFCOUNTER
\newcount\appcounter\appcounter=\the\APPCOUNTER
\newcount\PAGENO\PAGENO=\the\pageno
\newcount\FIGCOUNTER\FIGCOUNTER=0
\newcount\TABCOUNTER\TABCOUNTER=0
\newif\ifREFSYS\REFSYSfalse 
\newif\ifPHYSREVREF\PHYSREVREFtrue 
\newif\ifSETREFHEADING\SETREFHEADINGtrue
\newif\ifNOPAGENO\NOPAGENOfalse 
\newif\ifZPHYSREF\ZPHYSREFtrue 
\newif\ifCAMERAREADY\CAMERAREADYfalse   
\newif\ifTHESISFORMAT\THESISFORMATfalse   
\newif\ifNOCHAPFORMNUMBER\NOCHAPFORMNUMBERfalse 
\newif\ifFIRSTPRINTCONTENT\FIRSTPRINTCONTENTfalse
\newif\ifSETREFNUM 
\newif\ifSETREF 
\newwrite\REFFILE 
\newwrite\CONFILE 
%
\def\sqr#1#2{{\vcenter{\vbox{\hrule height.#2pt            
\hbox{\vrule width.#2pt height#1pt \kern#1pt
  \vrule width.#2pt}
\hrule height.#2pt}}}}

%
%
%
%
%
%
%
%
\def\thesisout{\output={\shipout\vbox{\nointerlineskip
        \vbox to 2truecm{\vskip 1.5truecm\ifnum\the\pageno>0
        \hbox{\hskip 2truecm\line{
        \ifodd\the\pageno
        \ifvoid\SECTIONBOX\ifvoid\CHAPTERBOX\else\copy\CHAPTERBOX\fi
        \else\topmark\fi\hfil\rm\the\pageno\else\rm\the\pageno\hfil
        \ifvoid\CHAPTERBOX\else\copy\CHAPTERBOX\fi\fi}}
        \fi\vfil}
        \hbox{\hbox to 2truecm{\hfill}
        \vtop to\vsize{\boxmaxdepth=\maxdepth\pagecontents\vfil}}}
        \global\advance\pageno by 1
        \ifnum\outputpenalty>-20000\else\dosupereject\fi}}
\def\filbreak{\ifdim\pagetotal >.75\pagegoal\ifdim\pagetotal <\pagegoal
        \NEWPAGE\fi\fi}
\def\paperout{\output={\shipout\vbox{\vbox to 1.5truecm{\vfill}
        \hbox{\hbox to 22truemm{\hfill}
        \vtop to\vsize{ \boxmaxdepth=\maxdepth\pagecontents\vfil}}
        \ifNOPAGENO\else
        \nointerlineskip\vbox to 3.2truecm{\vfil\vskip.5truecm
        \hbox{\hskip 22truemm\line{\centerline{\rm---\the\pageno---}}}
        \vfil}\fi}
        \global\advance\pageno by 1
        \ifnum\outputpenalty>-20000\else\dosupereject\fi}}
\def\filbreak{\ifdim\pagetotal >.75\pagegoal\ifdim\pagetotal <\pagegoal
        \NEWPAGE\fi\fi}
\def\NEWPAGE{\vfill\eject}
\def\BYE{\NEWPAGE\REFSYSOFF\end}
%
%
\def\TODAY{\ifcase\month\or January\or February\or March\or April\or May
        \or June\or July\or August\or September\or October\or
        November\or December\fi\space\number\day, \number\year}
\long\def\DATELINE#1{\setbox\datebox=\hbox to \hsize{
                \vtop{\hsize=0.5\hsize\hangindent=0pt\hangafter=-10\noindent#1}
                \hfil\TODAY }}
\def\TITLE#1{\titleskip=\bigskipamount
        \setbox\brolbox=\copy\titlebox
        \advance\titleskip by-\dp\brolbox
        \setbox\brolbox=\centerline{\fourteenbf#1}
        \advance\titleskip by-\ht\brolbox
        \setbox\titlebox=\vbox{\unvbox\titlebox
                                \vskip\titleskip\centerline{\fourteenbf#1}}}

\def\ADRES#1{\setbox\ADDRESSBOX=\vbox{\baselineskip 13truept
\centerline{\elevenit Instituut
voor Theoretische Natuurkunde en Materialen Studie Centrum}
\centerline{\elevenit Rijksuniversiteit Groningen, Nijenborgh 4}
\centerline{\elevenit NL-9747 AG Groningen, Nederland}
\centerline{\eightrm E-mail: #1}}}

\long\def\ABSTRACT#1{\setbox\abstractbox=\vbox{\hsize=5truein%
\noindent\eightrm\baselineskip=13truept#1}}
\long\def\FOOTNOTE#1{\setbox\footnotebox=\vbox{\vfil
        \hrule width 5truecm\parindent=0truemm\par\strut#1}}
\def\TITLEPAGE{\box\titlebox\bigskip
        \vbox{\hbox to\hsize{\hfil\vbox{\unvbox\AUTHORBOX}\hfil}}\smallskip
        \vbox{\hbox to\hsize{\hfil\vbox{\unvbox\ADDRESSBOX}\hfil}}\bigskip
        \ifvoid\abstractbox\else
        \medskip\centerline{\tenrm ABSTRACT}\medskip
        \centerline{\box\abstractbox}\noindent\fi}
%
%
\def\REFSYSON#1{\REFSYStrue\SETREFNUMtrue\SETREFtrue
        \xdef\REFSYSFIL{#1}\immediate\openout\REFFILE=\REFSYSFIL.REF
        \xdef\CONSYSFIL{#1}\immediate\openout\CONFILE=\CONSYSFIL.CON
        \global\pageno=\the\PAGENO\global\CHAPNUMBER=\the\chapnumber
        \global\APPCOUNTER=\the\appcounter\global\REFCOUNTER=\the\refcounter}
\def\REFSYSOFF{\ifREFSYS{
        \immediate\write\REFFILE{\string\PAGENO\string=\the\pageno}}
        \immediate\write\REFFILE{\string\chapnumber\string=\the\CHAPNUMBER}
        \immediate\write\REFFILE{\string\appcounter\string=\the\APPCOUNTER}
        \immediate\write\REFFILE{\string\refcounter\string=\the\REFCOUNTER}
        \immediate\closeout\REFFILE\immediate\closeout\CONFILE\fi
        \global\REFSYSfalse}
\def\PRINTREF#1{\SETREFtrue\ifREFSYS\global\REFCOUNTER=0\else
        \global\REFCOUNTER=\the\refcounter\fi\ifSETREFHEADING\HEADING{References}\fi
        \input #1.REF\SETREFHEADINGfalse}
\def\PRINTREFNOW{\bigskip\REFSYSOFF\input \REFSYSFIL.REF}

\def\REFA#1#2{\def\broldef{\string\NONUMBER}\def\droldef{\string#2}
        \ifx\droldef\broldef\vbox{\immediate\write\REFFILE{%
        \noexpand\ARTICLE#1{\string\NONUMBER}}}
        \else\global\advance\REFCOUNTER by1
        \ifSETREFNUM\the\REFCOUNTER\fi
        \vbox{\immediate\write\REFFILE{%
        \noexpand\ARTICLE#1{\def\string#2{\the\REFCOUNTER}}}}
        \xdef#2{\the\REFCOUNTER}\fi}
\long\def\ARTICLE #1/#2/#3/#4/#5/#6{
        \def\broldef{\string\NONUMBER}\def\droldef{\string#6}
        \ifSETREF\par\noindent\hangindent=24pt\hangafter=1
        \ifx\droldef\broldef\hbox to 24pt{\hfil}\else
        \global\advance\REFCOUNTER by1
        \hbox to 24pt{\hfil\the\REFCOUNTER.\ }\fi
        {\rm#1}\ifCAMERAREADY: \else, \fi{\sl#2}\ \ifPHYSREVREF{\bf#3}\else$
        \underline{\hbox{#3}}$\fi, #4 {(\rm#5).}\else
        \ifx\droldef\broldef\else#6\fi\fi}

\def\REFB#1#2{\vbox{\global\advance\REFCOUNTER by1
        \immediate\write\REFFILE{%
        \noexpand\BOOK#1{\string\def\string#2{\the\REFCOUNTER}}}
        \xdef#2{\the\REFCOUNTER}}\ifSETREFNUM\the\REFCOUNTER\fi}
\def\REFC#1#2{\vbox{\global\advance\REFCOUNTER by1
        \immediate\write\REFFILE{%
        \noexpand\PROCEEDING#1{\string\def\string#2{\the\REFCOUNTER}}}
        \xdef#2{\the\REFCOUNTER}}\ifSETREFNUM\the\REFCOUNTER\fi}
\def\REFD#1#2{\vbox{\global\advance\REFCOUNTER by1
        \immediate\write\REFFILE{%
        \noexpand\REFTEXT#1{\string\def\string#2{\the\REFCOUNTER}}}
        \xdef#2{\the\REFCOUNTER}}\ifSETREFNUM\the\REFCOUNTER\fi}
\def\REF#1{\ifZPHYSREF\ $\lbrack#1\rbrack$\else$^{#1}$\fi}
%
%
\long\def\BOOK #1/#2/#3/#4{\ifSETREF\global\advance\REFCOUNTER by1
        \par\noindent\hangindent=24pt\hangafter=1\hbox to 24pt{%
        \hfil\the\REFCOUNTER.\ }{\rm#1}\ifCAMERAREADY: in $
        \underline{\hbox{#2}}$, \else, {\sl#2}, \fi{(\rm#3).}\else#4\fi}
\long\def\PROCEEDING #1/#2/#3/#4/#5{\ifSETREF\global\advance\REFCOUNTER by1
        \par\noindent\hangindent=24pt\hangafter=1\hbox to 24pt{%
        \hfil\the\REFCOUNTER.\ }{\rm#1}\ifCAMERAREADY: in $
        \underline{\hbox{#2}}$, \else, {\sl#2}, \fi
        edited by #3, {\rm(#4).}\else#5\fi}
\long\def\REFTEXT #1/#2{\ifSETREF\global\advance\REFCOUNTER by1
        \par\noindent\hangindent=24pt\hangafter=1\hbox to 24pt{%
        \hfil\the\REFCOUNTER.\ }{\rm#1}\else#2\fi.}

\long\def\THEOREM#1#2{\setbox\brolbox=\hbox{#1}\broldimen=\wd\brolbox
        \par\noindent\hbox to \broldimen{\unhbox\brolbox}{#2}}

\long\def\ITEM#1#2{\global\advance\ITEMCOUNTER by1\global\ITEMITEMCOUNTER=0
        \setbox\brolbox=\hbox{\hbox to 24pt{\hfil\the\ITEMCOUNTER.\ }{#1}}
        \broldimen=\wd\brolbox\hbox{}
        \par\noindent\hangindent=\broldimen\hangafter=1
        \hbox to \broldimen{\unhbox\brolbox}{#2}}

\long\def\ITEMITEM#1#2{\setbox\brolbox=\hbox to 36pt{
\hfil\ifcase\ITEMITEMCOUNTER a\or b\or c\or d\or e\or f\or g\or h\fi.\ }
\hbox{}
\par\noindent\hangindent=36pt\hangafter=1\hbox to 36pt{\unhbox\brolbox}#1%
\par\noindent\line{\advance \hsize by -36pt\hbox to 36pt{\hfil}%
\vtop{\noindent#2}}\global\advance\ITEMITEMCOUNTER by1}
%
%

\def\leaderfill{\leaders\hbox to 1em{\hss.\hss}\hfill}
\def\CHAPTERCONTENT#1#2#3{\line{\hbox to 24pt{#1.\hfil}#2\leaderfill%
\hbox to 24pt{\hfil#3}}}
\def\SECTIONCONTENT#1#2#3#4{\line{\hbox to 24pt{}\hbox to 24pt{#1.#2\hfil}#3%
\leaderfill\hbox to 24pt{\hfil#4}}}
\def\APPENDIXCONTENT#1#2#3{\line{Appendix \ifcase#1 A\or B\or C\or D\fi.\ %
#2\leaderfill\hbox to 24pt{\hfil#3}}}
\def\HEADINGCONTENT#1#2{\line{#1\leaderfill\hbox to 24pt{\hfil#2}}}
%
%
\long\def\TAB#1{\vbox{\global\advance\TABCOUNTER by 1
        \setbox\brolbox=\hbox{Table.%
        \uppercase\expandafter{\romannumeral\the\TABCOUNTER}\ \ }
        \null\par\noindent\hangindent=\wd\brolbox\hangafter=1%
        \box\brolbox\null#1}}
\long\def\FIGKRIS#1{\vbox{\global\advance\FIGCOUNTER by 1
        \setbox\brolbox=\hbox{Fig.99\ }
        \null\par\noindent\hangindent=\wd\brolbox\hangafter=1%
        \hbox to \wd\brolbox{Fig.\the\FIGCOUNTER\hfil}{\rm#1}}}
%
%
\def\CHAPCOUNTER{\ifnum\APPCOUNTER < 0
{\the\CHAPNUMBER}\else{\ifcase\APPCOUNTER A\or B\or C\or D\fi}\fi}
\def\CHAPTER#1{\global\advance\CHAPNUMBER by 1\global\SECTNUMBER=0
        \setbox\CHAPTERBOX=\hbox{\bf \the\CHAPNUMBER.\ #1}
        \mark{\hbox{\unhcopy\CHAPTERBOX}}
        \setbox\SECTIONBOX=\hbox{\unhcopy\CHAPTERBOX}
        \ifNOCHAPFORMNUMBER\else\global\FORMCOUNTER=0\fi\medskip
        \ifREFSYS\edef\brolbrolbrol{\immediate\write\CONFILE{%
        \string\CHAPTERCONTENT{\the\CHAPNUMBER}{#1}{\noexpand\the\pageno}}}
        \brolbrolbrol\fi
        \line{\twelvebf\uppercase\expandafter{\the\CHAPNUMBER}.\ %
        {#1}\hfil}\nobreak\noindent}%
\def\SECTION#1{\global\advance\SECTNUMBER by 1\global\SECTNUMBERA=0
        \mark{\hbox{\bf \the\CHAPNUMBER.\the\SECTNUMBER.\ #1}}
        \setbox\SECTIONBOX=\hbox{\bf \the\CHAPNUMBER.\the\SECTNUMBER.\ #1}
        \ifREFSYS\edef\brolbrolbrol{\immediate\write\CONFILE{%
        \string\SECTIONCONTENT{\the\CHAPNUMBER}{\the\SECTNUMBER}{#1%
        }{\noexpand\the\pageno}}}\brolbrolbrol\fi
        \filbreak\medskip
        \line{\twelveit\uppercase\expandafter{\the\CHAPNUMBER}.%
        \the\SECTNUMBER\ #1\hfil}\nobreak\noindent}
\def\SECTIONA#1{\global\advance\SECTNUMBERA by 1\global\SECTNUMBERB=0
        \mark{\hbox{\bf
        \the\CHAPNUMBER.\the\SECTNUMBER.\the\SECTNUMBERA.\ #1}}
        \setbox\SECTIONBOX=\hbox{\bf
        \the\CHAPNUMBER.\the\SECTNUMBER.\the\SECTNUMBERA.\ #1}
        \ifREFSYS\edef\brolbrolbrol{\immediate\write\CONFILE{%
        \string\SECTIONCONTENT{\the\CHAPNUMBER}{\the\SECTNUMBER}{#1%
        }{\noexpand\the\pageno}}}\brolbrolbrol\fi
        \filbreak\medskip
        \line{\twelveit\uppercase\expandafter{\the\CHAPNUMBER}.%
        \the\SECTNUMBER.\the\SECTNUMBERA\ #1\hfil}\nobreak\noindent}
\def\HEADING#1{\ifREFSYS\edef\brolbrolbrol{\immediate\write\CONFILE{%
        \string\HEADINGCONTENT{#1}{\noexpand\the\pageno}}}
        \brolbrolbrol\fi\medskip
        \line{\twelvebf{#1}\hfil}\nobreak\noindent}%
\def\APPENDIX#1{\global\advance\APPCOUNTER by 1\global\SECTNUMBER=0%
        \global\FORMCOUNTER=0\global\CHAPNUMBER=0\filbreak\medskip
        \ifREFSYS\edef\brolbrolbrol{\immediate\write\CONFILE{%
        \string\APPENDIXCONTENT{\the\APPCOUNTER}{#1}{\noexpand\the\pageno}}}
        \brolbrolbrol\fi
        \line{\twelvebf{Appendix\ %
        \ifcase\APPCOUNTER A\or B\or C\or D\or E\or F\fi:\ #1}%
        \rm\hfil}\nobreak\medskip}
%
%
\def\FORMNUMBER{\ifNOCHAPFORMNUMBER(\the\FORMCOUNTER)\else
                (\CHAPCOUNTER.\the\FORMCOUNTER)\fi}
\long\def\FORM#1#2 \par{\global\advance\FORMCOUNTER by1
        $$#1\quad#2\eqno\FORMNUMBER$$\par\noindent}
\def\CR#1{\quad#1&\EQANUMBER\cr\noalign{\global\advance\EQACOUNTER by1}}
\def\CRZ#1{\quad#1&\FORMNUMBER\cr}      
\def\CRY#1{\quad#1&\FORMNUMBER\cr\noalign{\global\advance\FORMCOUNTER by1}}
\def\EQANUMBER{\ifNOCHAPFORMNUMBER(\the\FORMCOUNTER\EQALETTER)\else
                (\CHAPCOUNTER.\the\FORMCOUNTER\EQALETTER)\fi}
\def\EQALETTER{\ifcase\EQACOUNTER a\or b\or c\or d\or e\or f\or g\or h\fi}
\def\NOALIGN#1{\noalign{\smallskip}\noalign{\vbox{\noindent#1\hfil}}
                \noalign{\smallskip}}
\long\def\EQA#1 \par{\global\advance\FORMCOUNTER by1
                        \global\EQACOUNTER=0$$\eqalignno{#1}$$\par\noindent}
\def\CUR#1/{{\global\advance\FORMCOUNTER by #1\FORMNUMBER\global\advance
        \FORMCOUNTER by -#1}}
\def\CURE#1#2/{{\global\advance\FORMCOUNTER by #1%
\ifNOCHAPFORMNUMBER(\the\FORMCOUNTER#2)%
\else(\CHAPCOUNTER.\the\FORMCOUNTER#2)\fi%
\global\advance\FORMCOUNTER by -#1}}
%
%
\voffset=-2truecm\hoffset=-2truecm 
\hsize=152truemm\vsize=218truemm\rm
\normalbaselineskip=15truept plus0pt minus0pt\normalbaselines
\normparindent=5truemm\parindent=6truemm\parskip=0 truemm plus0pt
\belowdisplayshortskip=3 truemm plus1pt minus1pt
\belowdisplayskip=3 truemm plus1pt minus1pt
\abovedisplayshortskip=0 truemm
\abovedisplayskip=0 truemm
\smallskipamount=5 truept plus0pt minus0pt
\medskipamount=15 truept plus0pt minus0pt
\bigskipamount=20 truept plus0pt minus0pt
\paperout               
\NOCHAPFORMNUMBERtrue
%
%
\hyphenation{quad-ratic di-men-sion-al for-mula prod-uct equi-li-bri-um
              tri-di-ag-on-al tri-di-ag-onal-iz-at-ion  non-equi-li-bri-um}

\def\ORDER#1{\hbox{${\cal O}(#1)$}}
\def\-{\discretionary{-}{}{-}}
\def\KET#1{\hbox{$|#1\rangle$}}
\def\BRA#1{\hbox{$\langle#1|$}}
\def\BRACKET#1#2{\hbox{$\langle#1|#2\rangle$}}
\def\EXPECT#1{\hbox{$\langle#1\rangle$}}
\def\NEXTFIG{{\global\advance\FIGCOUNTER by 1}}
\def\FIG{\the\FIGCOUNTER.\ }
\def\CURFIG#1/{{\global\advance\FIGCOUNTER by #1}\the\FIGCOUNTER%
{\global\advance\FIGCOUNTER by -#1}}
\input epsf
\ZPHYSREFtrue
\REFSYSON{TEXREF}
\font\eightrm=cmr8
\font\eightsl=cmsl8
\font\eightit=cmmi8

\def\KLEIN#1{\baselineskip=13truept plus 0pt minus 0pt
{\eightrm%
\textfont0=\eightrm\scriptfont0=\scriptscriptfont0%
\textfont1=\eightit\scriptfont1=\scriptscriptfont1#1}}
\def\BAR{\bar }
\def\NOBAR{}

\line{To appear in Computer Physics Communications\hfil\TODAY}
\bigskip
\centerline{\fourteenbf Quantum Computer Emulator}
\bigskip
%
\vbox{\centerline{\tenrm Hans De Raedt$^1$, Anthony H. Hams$^1$,
and Kristel Michielsen$^2$}}
\smallskip
\vbox{\baselineskip 13truept
\centerline{\elevenit $^{1}$Institute for Theoretical Physics and Materials Science Centre}
\centerline{\elevenit $^{2}$Laboratory for Biophysical Chemistry}
\centerline{\elevenit University of Groningen, Nijenborgh 4}
\centerline{\elevenit NL-9747 AG Groningen, The Netherlands}
\centerline{\eightrm E-mail: deraedt@phys.rug.nl,
A.H.Hams@phys.rug.nl
}
\centerline{\eightrm E-mail: K.F.L.Michielsen@chem.rug.nl}
\centerline{\eightrm http://rugth30.phys.rug.nl/compphys}
}
\hbox{}\smallskip
%
\vbox{\centerline{\tenrm Koen De Raedt}}
\smallskip
\vbox{\baselineskip 13truept
\centerline{\elevenit \phantom{$^{1}$}European Marketing Support, Vlasakker 21}
\centerline{\elevenit B-2160 Wommelgem, Belgium}
}
\hbox{}\bigskip
\centerline{\vbox{\eightrm\hsize=140truemm\noindent%
We describe a quantum computer emulator for a generic, general purpose
quantum computer. This emulator consists of
a simulator of the physical realization of the quantum computer and
a graphical user interface to program and control the simulator.
We illustrate the use of the quantum computer emulator through various
implementations of the Deutsch-Jozsa and Grover's database search
algorithm.
}}
\medskip
\centerline{\vbox{\eightrm\hsize=140truemm\noindent%
}}
\hbox{}\smallskip

\CHAPTER{Introduction}

Recent progress in the field of quantum information processing has
opened new prospects to use quantum mechanical phenomena
for processing information.
The operation of elementary quantum logic gates
using ion traps, cavity QED, and NMR technology has been demonstrated.
A primitive Quantum Computer (QC)\REF{
\REFD{J.A. Jones, and M. Mosca,
``Implementation of a quantum algorithm on a nuclear magnetic resonance
quantum computer'',
J. Chem. Phys. {\bf109}, 1648 - 1653 (1998)/}{\JONESone}-
{\SETREFNUMfalse
\REFD{J.A. Jones, M. Mosca, and R.H. Hansen,
``Implementation of a quantum search algorithm on a quantum computer'',
Nature (London) {\bf393}, 344 - 346 (1998)/}{\JONEStwo}
\REFD{I.L. Chuang, L.M.K. Vandersypen, Xinlan Zhou, D.W. Leung, and S. Lloyd,
``Experimental realization of a quantum algorithm'',
Nature {\bf393}, 143 - 146 (1998)/}{\CHUANGtwo}
}
\REFD{I.L. Chuang, N. Gershenfeld, and M. Kubinec,
``Experimental implementation of Fast Quantum Searching'',
Phys. Rev. Lett. {\bf80}, 3408 - 3411 (1998)/}{\CHUANGthree}
}
and secure quantum cryptographic systems have been build\REF{
\REFD{C. Bennett, and G. Brassard,
``The Dawn of a New Era for Quantum Cryptography: The Experimental
Prototype is Working'',
SIGACT News {\bf 20}, 78 - 82 (1989)/}{\BENNETTone}-
{\SETREFNUMfalse
\REFD{C. Bennett, F. Bessette, G. Brassard, L.G. Salvail, and J. Smolin,
``Experimental Quantum Cryptography'',
J. Cryptology {\bf5}, 3 - 28 (1992)/}{\BENNETTtwo}
}
\REFD{A. Ekert, J. Rarity, P. Tapster, and G. Palma,
``Practical Quantum Cryptography Based on Two\-Photon Interferometry'',
Phys. Rev. Lett. {\bf69}, 1293 - 1295 (1992)/}{\EKERTtwo}
}.
Recent theoretical work has shown that a QC
has the potential of solving certain computationally hard
problems such as factoring integers
and searching databases much faster than a conventional computer\REF{
\REFD{P. Shor,
``Algorithms for quantum computation: Discrete logarithms and factoring'',
in {{\sl Proc. 35th Annu. Symp. Foundations of Computer Science}},
S. Goldwasser ed., 124 (IEEE Computer Soc., Los Alamitos CA, 1994)/}{\SHORone}-
{\SETREFNUMfalse
\REFD{I.L. Chuang, R. Laflamme, P.W. Shor, and W.H. Zurek,
`` Quantum Computers, Factoring and Decoherence'',
Science {\bf230}, 1663 - 1665 (1995)/}{\CHUANGone}
\REFD{A.Yu. Kitaev,
``Quantum measurements and the Abelian stabiliser problem'',
e-print {quant-ph/9511026}/}{\KITAEV}
\REFD{L.K. Grover,
``A fast quantum mechanical algorithm for database search'',
in {\sl Proc. of the 28th Annual ACM Symposium of Theory of Computing}
(ACM, Philadelphia 1996)/}{\GROVERzero}
\REFD{L.K. Grover,
``Quantum Computers can search arbitrary large databases by a single query'',
Phys. Rev. Lett. {\bf 79}, 4709 - 4712 (1997)/}{\GROVERone}
}
\REFD{L.K. Grover,
``Quantum Computers can search rapidly by using almost any transformation'',
Phys. Rev. Lett. {\bf 80}, 4329 - 4332 (1998)/}{\GROVERtwo}
}.

The fact that a QC might be more powerful than an ordinary computer is based
on the notion that a quantum system can be in any superposition of states
and that interference of these states allows exponentially
many computations to be done in parallel\REF{
\REFD{D. Aharonov,
``Quantum Computation'',
e-print {quant-ph/9812037}/}{\AHARONOVone}
}.
This intrinsic parallelism might be used to solve other
difficult problems as well, such as for example
the calculation of the physical properties of quantum many\-body systems\REF{
\REFD{N.J. Cerf, and S.E. Koonin,
``Monte Carlo simulation of quantum computation'',
Mathematics and Computers in Simulation {\bf 47}, 143 - 152 (1998)/}{\CERFone}-
{\SETREFNUMfalse
\REFD{C. Zalka,
``Simulating quantum systems on a quantum computer'',
Proc. R. Soc. London {\bf A454}, 313 - 322 (1998)/}{\ZALKAone}
\REFD{B.M. Terhal, and S.P. DiVincenzo,
``On the problem of equilibration and the computation of correlation
functions on a quantum computer'',
e-print {quant-ph/9810063}/}{\TERHALone}
}
\REFD{H. De Raedt, A.H. Hams, K. Michielsen, S. Miyashita, and K. Saito,
``Quantum Statistical Mechanics on a Quantum Computer'',
Prog. Theor. Phys. (in press), e-print {quant-ph/9911037}/}{\QCzero}
}.
In fact, part of Feynman's original motivation
to consider QC's was that they might be used
as a vehicle to perform exact simulations of quantum mechanical phenomena\REF{
\REFD{R.P. Feynman,
``Simulating Physics with Computers'',
Int. J. Theor. Phys. {\bf21}, 467 - 488 (1982)/}{\FEYNMAN}}.

Just as simulation is an integral part
of the design process of each new generation of microprocessors,
software to emulate the physical model representing the hardware implementation
of a quantum processor may prove essential.
In contrast to conventional digital
circuits (which may be build using vacuum tubes, relays, CMOS etc.)
where the internal working of each basic unit is
irrelevant for the logical operation of the whole machine
(but extremely relevant for the speed of operation and the cost
of the machine of course),
in a QC the internal quantum dynamics of each elementary constituent
is a key ingredient of the QC itself.
Therefore it is essential to incorporate into a simulation model, the
physics of the elementary units that make up the QC.

Theoretical work on quantum computation usually assumes the existence
of units that perform highly idealized unitary operations.
However, in practice these operations are difficult to realize:
Disregarding decoherence,
a hardware implementation of a QC will perform
unitary operations that are more complicated than those considered
in most theoretical work. Therefore it is important to have theoretical
tools to validate designs of physically realizable quantum processors.

This paper describes a Quantum Computer Emulator (QCE)
to emulate various hardware designs of QC's.
The QCE simulates the physical processes
that govern the operation of the hardware quantum processor,
strictly according to the laws of quantum mechanics.
The QCE also provides an environment to debug and execute quantum algorithms (QA's)
under realistic experimental conditions.
This is illustrated for several
implementations of the Deutsch\-Jozsa\REF{
\REFD{D. Deutsch, and R. Jozsa,
``Rapid solution of problems by quantum computation'',
Proc. R. Soc. Lond. A{\bf439}, 553 - 558 (1992)/}{\DEUTSCHJOZSA},
\REFD{D. Collins, K.W. Kim, and W.C. Holton,
``Deutsch-Jozsa algorithm as a test of quantum computation'',
Phys. Rev. A {\bf58}, R1633 - R1636 (1998)/}{\COLLINSone}
}
and Grover's database search algorithm\REF{\GROVERone,\GROVERtwo}
on QC's using ideal and more realistic units,
such as those used in the 2-qubit NMR QC\REF{\CHUANGtwo,\CHUANGthree}.
Elsewhere\REF{\QCzero} we present results
of a QA to compute
the thermodynamic properties of quantum many-body systems
obtained on a 21-qubit hard-coded version of the QCE.
The QCE software runs in a W98/NT4 environment and may be
dowloaded from {\sl http://rugth30.phys.rug.nl/compphys/qce.htm}.

\CHAPTER{QCE: Quantum Computer Emulator}

Generically, hardware QC's are modeled
in terms of quantum spins (qubits) that
evolve in time according to the time-dependent Schr\"odinger
equation (TDSE)

\FORM{
i{\partial \over\partial t} \KET{\Phi(t)}= H(t) \KET{\Phi(t)}
},

\edef\TDSE{\immediate\CUR{0}/}%
in units such that $\hbar=1$ and where

\EQA{\KET{\Phi(t)}=
a(\downarrow,\downarrow,\ldots,\downarrow;t)
\KET{\downarrow,\downarrow,\ldots,\downarrow}
&+a(\uparrow,\downarrow,\ldots,\downarrow;t)
\KET{\uparrow,\downarrow,\ldots,\downarrow}+ \ldots \cr
&+a(\uparrow,\uparrow,\ldots,\uparrow;t)
\KET{\uparrow,\uparrow,\ldots,\uparrow}
\CRZ,}

describes the state of the whole QC at time $t$.
The complex coefficients
$a(\downarrow,\downarrow,\ldots,\downarrow;t),\ldots,
a(\uparrow,\uparrow,\ldots,\uparrow;t)$ completely specify the
state of the quantum system.
The time-dependent Hamiltonian $H(t)$ takes the form\REF{
\REFD{We use a notation that closely resembles the symbols
in the QCE's graphical user interface/}{\XXXXXX}}

\EQA{
H(t)=&-\sum_{j,k=1}^L\sum_{\alpha=x,y,z} J_{j,k,\alpha}(t) S_j^\alpha S_k^\alpha\cr
&-\sum_{j=1}^L\sum_{\alpha=x,y,z}
\left( h_{j,\alpha,0}(t) + h_{j,\alpha,1}(t) \sin (f_{j,\alpha} t + \varphi_{j,\alpha})
\right) S_j^\alpha
\CRZ,}

\edef\FULLHAM{\immediate\CUR{0}/}%
where the first sum runs over all pairs $P$ of spins (qubits),
$S_j^\alpha$ denotes the $\alpha$\-th component of the spin-1/2
operator representing the $j$-th qubit,
$J_{j,k,\alpha}(t) $ determines the strength of the interaction between
the qubits labeled $j$ and $k$,
$h_{j,\alpha,0}(t)$ and $h_{j,\alpha,1}(t)$ are
the static (magnetic) and periodic (RF) field acting on the $j$-th spin respectively.
The frequency and phase of the periodic field are denoted by $f_{j,\alpha}$
and $\varphi_{j,\alpha}$.
The number of qubits is $L$ and the dimension of the Hilbert space $D=2^L$.

Hamiltonian \CUR{0}/ is sufficiently
general to capture the salient features of most physical models of QC's.
Interactions between qubits that involve different spin components
have been left out in \CUR{0}/ because we are not aware of a candidate
technology of QC where these would be important. Incorporating these
interactions requires some trivial additions to the QCE program.

A QA for QC model \FULLHAM\ consists of a sequence of
elementary operations which we will call micro instructions (MI's) in the sequel.
They are not exactly playing the same role as MI's do in
digital processors, they merely represent the smallest units of operation
the quantum processor can carry out.
The action of a MI on the state \KET{\Psi}
of the quantum processor is defined by specifying
how long it acts (i.e. the time interval it is active),
and the values of all the $J$'s and $h$'s appearing in \FULLHAM.
The $J$'s and $h$'s are fixed during the operation of the MI.
A MI transforms the input state \KET{\Psi(t)}
into the output state \KET{\Psi(t+\tau)} where $\tau$ denotes the
time interval during which the MI is active.
During this time interval the only time-dependence
of $H(t)$ is through the sinusoidal modulation of the fields on the spins.

Procedures to construct unconditionally stable,
accurate and efficient algorithms to solve the TDSE
of a wide variety of continuum and lattice models
have been reviewed elsewhere\REF{
\REFD{H. De Raedt,
``Product Formula Algorithms for Solving the Time\-Dependent Schr\"odinger Equation'',
Comp. Phys. Rep. {\bf7}, 1 - 72 (1987)/}{\HDRCPR}-
{\SETREFNUMfalse
\REFD{H. De Raedt and K. Michielsen,
``Algorithm to Solve the Time\-Dependent
Schr\"odinger Equation for a Charged Particle in an Inhomogeneous Magnetic
Field: Application to the Aharonov\-Bohm Effect'',
Comp. in Phys. {\bf8}, 600 - 607 (1994)/}{\HDRKRMone}
\REFD{H. De Raedt,
``Quantum Dynamics in Nano-Scale Devices'',
Computational Physics, 209 -- 224, ed. K.-H. Hoffmann and
M. Schreiber, Springer 1996/}{\HDRHOFFMANN}
}
\REFD{H. De Raedt, ``Computer Simulation of Quantum Phenomena
in Nano-Scale Devices'', 107 -- 146, Annual Reviews of Computational Physics IV,
ed. D. Stauffer, World Scientific 1996/}{\HDRSTAUFFER}
}.
A detailed account of the application of this approach
to two\-dimensional quantum spin models can be found in\REF{
\REFD{P. de Vries, and H. De Raedt
{``Solution of the time-dependent Schr\"odinger equation for two-dimensional
spin-1/2 Heisenberg systems''},
Phys. Rev. B {\bf47}, 7929 - 7937 (1993)/}{\PEDROone}
}.
Here we limit ourselves to a discussion of the basic steps in the
construction of an algorithm to solve the TDSE for a arbitrary
model of the type \FULLHAM.
According to \CUR{-1}/ the time evolution of the QC,
i.e. the solution of TDSE \TDSE, is determined by the unitary transformation
$U(t+\tau,t)\equiv\exp_{+}(-i\int_{t}^{t+\tau} H(u) du)$,
where $\exp_{+}$ denotes the time-ordered exponential function.
Using the semi\-group property of $U(t+\tau,t)$ we can write

\FORM{U(t+\tau,t)=U(t+m\delta,t+(m-1)\delta)
\cdots U(t+2\delta,t+\delta)U(t+\delta,t)},

where $\tau=m\delta$ ($m\ge1$).
In general the first step is to replace each $U(t+(n+1)\delta,t+n\delta)$
by a symmetrized Suzuki product\-formula approximation\REF{\HDRCPR,
{\SETREFNUMfalse
\REFD{J. Huyghebaert, and H. De Raedt,
``Product formula methods for time\-depen\-dent Schr\"odinger problems'',
J. Phys. A:Math. Gen. {\bf23}, 5777 (1990)/}{\XXXXXX}
}
\REFD{M. Suzuki,
``General Decomposition Theory of Ordered Exponentials'',
Proc. Japan Acad. {\bf69}, Ser. B 161 - 166 (1993)/}{\XXXXXX}
}.
For the case at hand a convenient choice is
(other decompositions\REF{\PEDROone,
\REFD{M. Suzuki, S. Miyashita, and A. Kuroda,
``Monte Carlo Simulation of Quantum Spin Systems. I'',
Prog. Theor. Phys. {\bf58}, 1377 - 1387 (1977)/}{\SUZUKItwo}
}
work equally well but are somewhat less efficient for our purposes):

\EQA{U(t+(n+1)\delta,t+n\delta)\approx&\,{\widetilde U(t+(n+1)\delta,t+n\delta)}\CR,
\NOALIGN{}
{\widetilde U(t+(n+1)\delta,t+n\delta)}=&\,
e^{-i\delta H_z(t+(n+1/2)\delta)/2}
e^{-i\delta H_y(t+(n+1/2)\delta)/2}\cr
&\,\times e^{-i\delta H_x(t+(n+1/2)\delta)}
e^{-i\delta H_y(t+(n+1/2)\delta)/2}\cr
&\,\times e^{-i\delta H_z(t+(n+1/2)\delta)/2}
\CR,}

where

\EQA{H_\alpha(t)=&-\sum_{j,k=1}^L J_{i,j,\alpha} S_j^\alpha S_k^\alpha\cr
&-\sum_{j=1}^L
\left( h_{j,\alpha,0} + h_{j,\alpha,1} \sin( f_{j,\alpha} t + \varphi_{j,\alpha})
\right) S_j^\alpha
\quad;\quad\alpha=x,y,z\CRZ.
}

Note that in \CUR{0}/ we have omitted the time dependence of the
$J$'s and the $h$'s to emphasize that these parameters are fixed
during the execution of a particular MI.

Evidently ${\widetilde U(t+\tau,t)}$ is unitary by construction,
implying that the algorithm to solve the TDSE is unconditionally stable\REF{\HDRCPR}.
It can be shown that
$\Vert U(t+\tau,t) - {\widetilde U(t+\tau,t)}\Vert \le c \delta^3$, implying
that the algorithm is correct to second order in the time-step $\delta$\REF{\HDRCPR}.
If necessary, ${\widetilde U(t+\tau,t)}$
can be used as a building block to construct higher\-order algorithms\REF{
\REFD{H. De Raedt, and B. De Raedt,
``Applications of the Generalized Trotter Formula'',
Phys. Rev. A {\bf28}, 3575 - 3580 (1983)/}{\HDRone}-
{\SETREFNUMfalse
\REFD{M. Suzuki,
``Decomposition Formulas of Exponential Operators and Lie
Exponentials with some Applications to Quantum Mechanics
and Statistical Physics''
J. Math. Phys. {\bf26}, 601 - 612 (1985)/}{\SUZUKItwo}
}
\REFD{M. Suzuki,
``General Nonsymetric Higher-Order Decomposition of
Exponential Operators and Symplectic Integrators''
J. Phys. Soc. Jpn. {\bf61}, 3015 - 3019 (1995)/}{\SUZUKIthree}
}.
In practice it is easy to find reasonable values of $m$ such that
the results obtained no longer depend on $m$ (and $\delta$).
Then, for all practical purposes, these results
are indistinghuisable from the exact solution of the TDSE \TDSE.

As already indicated above, as basis states $\{\KET{\phi_n}\}$ we take
the direct product of the eigenvectors of the
$S_j^z$ (i.e. spin\-up $\KET{\uparrow}_j$ and spin\-down $\KET{\downarrow}_j$).
In this basis $e^{-i\delta H_z(t+(n+1/2)\delta)/2}$ changes the input state by altering
the phase of each of the basis vectors.
As $H_z$ is a sum of pair interactions it is trivial to rewrite this operation
as a direct product of 4x4 diagonal matrices (containing
the interaction\-controlled phase shifts) and 4x4 unit matrices.
Hence the computation of $\exp(-i\delta H_z(t+(n+1/2)\delta)/2)\KET{\Psi}$ has been
reduced to the multiplication of two vectors, element-by-element.
The QCE carries out \ORDER{P2^L} operations to perform this calculation
but a real QC operates on all qubits simultaneously
and would therefore only need \ORDER{P} operations.

Still working in the same representation, the action of
$e^{-i\delta H_y(t+(n+1/2)\delta)/2}$
can be written in a similar manner but the matrices that contain the
interaction\-controlled phase\-shift have to be replaced by
non-diagonal matrices. Although this does not present a real problem it is
more efficient and systematic to proceed as follows.
Let us denote by ${\cal X}$ (${\cal Y}$) the rotation by $\pi/2$ of all spins
about the $x$($y$)-axis. As

\EQA{
e^{-i\delta H_y(t+(n+1/2)\delta)/2}=&
{\cal X}{\cal X}^\dagger e^{-i\delta H_y(t+(n+1/2)\delta)/2}{\cal X}
{\cal X}^\dagger\cr
=&{\cal X} e^{-i\delta H_z^\prime(t+(n+1/2)\delta)/2}{\cal X}^\dagger
\CRZ,}

it is clear that the action of
$e^{-iw\delta H_y(t+(n+1/2)\delta)/2}$ can be computed by
applying to each qubit, the inverse of ${\cal X}$
followed by an interaction-controlled phase-shift and ${\cal X}$.
The prime in \CUR{0}/ indicates that
$J_{i,j,z}$, $h_{i,z,0}$, $h_{i,z,1}$ and $f_{i,z}$
in $H_z(t+(n+1/2)\delta)$
have to be replaced by
$J_{i,j,y}$, $h_{i,y,0}$, $h_{i,y,1}$ and $f_{i,y}$ respectively.
A similar procedure is used to compute the action of
$e^{-i\delta H_x(t+(n+1/2)\delta)}$: We only have to replace ${\cal X}$ by ${\cal Y}$.
The operation counts for $e^{-i\delta H_x(t+(n+1/2)\delta)}$
(or $e^{-i\delta H_y(t+(n+1/2)\delta)/2}$)
are \ORDER{(P+2) 2^L} and \ORDER{P+2} for the QCE and QC respectively.
On a QC the total operation count per time-step is \ORDER{3P+4}.

The operation count of the algorithm described above (and variations of it,
see e.g.\REF{\PEDROone}) increases linearly with
the dimension $D$ of the Hilbert space,
and cannot be improved in that sense (although there is may be room
for reducing the prefactor by more clever programming).
On the other hand one might be tempted to think that
for small $D$ the cost of an exact
diagonalization of the time\-dependent Hamiltonian \FULLHAM\
at each time\-step $\tau$ may be compensated for by absence of intermediate
time\-steps $\delta$. For the problem at hand this is unlikely
to be the case: The sinusoidal terms in \FULLHAM\ require the use
of a time\-step that is much smaller than the time\-step that
guarantuees a high accuracy of the Suzuki product formula. Therefore
in practice for time\-dependent Hamiltonian \FULLHAM, $\delta=\tau$
so that disregarding implementation issues,
the running time of the algorithm is very close to the theoretical limit.

\CHAPTER{Graphical User Interface}

The QCE consists of a QC simulator, described above, and a graphical user interface (GUI)
that controls the former.
The GUI considerably simplifies the task of specifying the MI's (i.e.
to model the hardware) and to execute quantum programs (QP's).
The QCE runs in a Windows 98/NT environment.
Using the GUI is very much like working with any other standard
MS\-Windows application.
The maximum number of qubits in the version of the QCE that is available
for distribution is limited to eight.
The QCE is distributed as a self-installing executable, containing
the program, documentation, and all the QP's discussed in this paper.
These QP's also illustrate the use of the GUI itself.

Some of the salient features of the GUI of the QCE are shown in figs.1-4.
The main window contains a window that shows the set of
MI's that is currently active and several
other windows (limited to 10) that contain QPs.
Help on a button appears when the mouse moves over the button,
a standard Windows feature.

Writing a QA on the QCE from scratch is a two-step process.
First one has to specify the MI's, taking into account
the particular physical realization of the QC that one wants
to emulate. The "MI" window offers all necessary tools to edit (see Fig.2)
and manipulate (groups of) MI's.
The second step, writing a QP, consists
of dragging and dropping MI's onto a ``QP'' window.

Each MI set has two reserved MI's: A break point (allowing
the QP to pause at a specified point) and a MI to initialize the QC.
Normally the latter is the first instruction in a QP.
Each QP window has a few buttons to control the operation of the QC.

The results of executing a QP appear in color-coded form at the bottom
of the corresponding program window. For each qubit the expectation
value of the three spin components are shown:
$Q^\alpha_j \equiv 1/2 - \EXPECT{S_j^\alpha}$ ($\alpha=x,y,z$)
green corresponds to 0, red to 1.
Usually only one row of values (the $z$\-component) will be of interest.
Optionally the QCE will generate text files with the numerical results
for further processing.

The QCE supports the use of QP's as MI's (see Figs.3,4).
QP's can be added to a particular MI set through the button labeled ``QP''.
During execution, a QP that is called from another QP will call
either another QP or a genuine MI from the currently loaded set of MI's.
The QCE will skip all initialization MI's except for the first one.
This facilitates the testing of QP that are used as sub\-QP's.
A QP calling a MI that cannot be found in the current MI set will
generate an error message and stop.

\CHAPTER{Applications}

Our aim is to illustrate how to use the QCE to simulate the QC
implemented using NMR techniques\REF{\JONESone,\JONEStwo,\CHUANGtwo,\CHUANGthree}.
A classical coin has been used to decide which of the two realizations
(i.e.\REF{\JONESone,\JONEStwo} or\REF{\CHUANGtwo,\CHUANGthree})
to take as an example.
In the NMR experiments\REF{\CHUANGtwo,\CHUANGthree}
the two nuclear spins of the ($^1$H and $^{13}$C atoms in
a carbon\-13 labeled chloroform)
molecule are placed in a strong static magnetic field in the
$+z$ direction. In the absence of interactions with other
degrees of freedom this spin-1/2 system can be modeled
by the hamiltonian

\FORM{H =- J_{1,2,z} S_1^z S_2^z
- h_{1,z,0} S_1^z
- h_{2,z,0} S_2^z
},

\edef\HAM{\immediate\CUR{0}/}%
where $h_{1,z,0}/2\pi\approx 500 \hbox{MHz}$,
$h_{2,z,0}/2\pi\approx125 \hbox{MHz}$, and
$J_{1,2,z}/2\pi\approx-215 \hbox{Hz}$\REF{\CHUANGtwo}.
It is amusing to note that the most simple spin-1/2 system, i.e. the Ising model,
can be used for quantum computing\REF{
\REFD{S. Lloyd,
``A potentially Realizable Quantum Computer,
{\sl Science} {\bf 261}, 1569 (1993)/}{\LLOYDone}-
{\SETREFNUMfalse
\REFD{G.P. Berman, G.D. Doolen, D.D. Holm, and V.I. Tsifrinovich,
``Quantum Computer on a class of one-dimensional Ising systems'',
Phys. Lett. {\bf A193}, 444 - 450 (1994)/}{\BERMANone}
\REFD{G.P. Berman, G.D. Doolen, R. Mainieri, and V.I. Tsifrinovich,
{\sl Introduction to Quantum Computers},
(World Scientific, 1998)/}{\BERMANzero}
\REFD{G.P. Berman, G.D. Doolen, and V.I. Tsifrinovich,
``Quantum Computation as a Dynamical Process'',
e-print {quant-ph/9904105}/}{\BERMANtwo}
}
\REFD{G.P. Berman, G.D. Doolen, G.V. L\'opez, and V.I. Tsifrinovich,
``Quantum Computation as a Dynamical Process'',
e-print {quant-ph/9909027}/}{\BERMANthree}
}

In the chloroform molecule
the antiferromagnetic interaction between the spins is much weaker
than the coupling to the external field and \CUR{0}/
is a diagonal matrix with respect to the basis states chosen, the ground state
of \CUR{0}/ is the state with the two spins up.
Following\REF{\CHUANGtwo} we denote this state
by $\KET{00}=\KET{0}\otimes\KET{0}=\KET{\uparrow\uparrow}$,
i.e. the state with spin up corresponds to a qubit \KET{0}.
A state of the $N$\-qubit QC will be denoted by
$\KET{x_1x_2\ldots x_N}=\KET{x_1}\otimes\KET{x_2}\ldots\KET{x_N}$.

It is expedient to write the TDSE for this problem
in frames of reference rotating with the nuclear spin.
Formally this is accomplished by substituting in \TDSE

\FORM{\KET{\Phi(t)}=e^{it(h_{1,z,0} S^z_1+h_{2,z,0} S^z_2)}\KET{\Psi(t)}
},

so that in the absence of RF-fields the time evolution of
$\Psi(t)$ is governed by the hamiltonian
$H =- J_{1,2,z} S_1^z S_2^z$.

This transformation removes from the sequence of elementary operations,
phase factors that are irrelevant for the value of the qubits.
Indeed, as the expectation value of a qubit is related to
the expectation value of the $z$ component of the spin:

\EQA{
Q_j\equiv Q^z_j=&{1\over2}- \BRACKET{\Phi(t)}{S_i^z|\Phi(t)}\CR,
\NOALIGN{and}
\BRACKET{\Phi(t)}{S_j^z|\Phi(t)}=&
\BRACKET{\Psi(t)}{e^{-it(h_{1,z,0} S^z_1+h_{2,z,0} S^z_2)}
S_j^z
e^{it(h_{1,z,0} S^z_1+h_{2,z,0} S^z_2)}|\Psi(t)}\cr
=&
\BRACKET{\Psi(t)}{S_j^z|\Psi(t)}\CR.
}

>From \CUR{0}/ it is clear that transformation \CUR{-1}/ has no net effect.
This is not the case for
the expectation values of the $x$ or $y$ component of the spins:
The phase factors induce an oscillatory behavior, reflecting the fact
that the spins are rotating about the $z$\-axis (see $(A.17$) for an example).
In the following it is implicitly assumed
that the basis states of the spins refer to
states in the corresponding rotating frame.

We now discuss the implementation on the QCE of two QA's
that have been tested
on an NMR QC\REF{\JONESone,\JONEStwo,\CHUANGtwo,\CHUANGthree}.

\SECTION{Deutsch-Jozsa algorithm}

This QA\REF{\DEUTSCHJOZSA,
\REFD{R. Cleve, A. Ekert, C. Macciavello, and M. Mosca,
``Quantum algorithms revisited'',
Proc. R. Soc. Lond. A {\bf 454}, 339 - 354 (1998)/}{\CLEVEone}
}
and its refinement\REF{\COLLINSone}
provide illustrative examples of how the intrinsic parallelism of a QC
can be exploited to solve certain decision problems.

Consider a function $f=f(x_1,x_2,\ldots,x_N)=0,1$ that transforms
the $N$ bits  $\{x_n=0,1\}$ to one output bit.
There are three classes of functions $f$: {\sl Constant} functions, returning
0 or 1 independent of the input $\{x_n\}$, {\sl balanced} functions that
give $f=0$ for exactly half of the $2^N$ possible inputs and
$f=1$ for the remaining inputs, and {\sl other} functions
that do not belong to one of the two other classes.
Some examples of {\sl constant} and {\sl balanced} functions are given in
tables 1 and 2.

The Deutsch-Jozsa (D\-J) algorithm
allows a QC to decide whether a function
is {\sl constant} or {\sl balanced}, given the additional piece
of information that functions of the type {\sl other} will not
be fed into the QC.
For a function of one input variable the D\-J problem is
equivalent to the problem of deciding if a coin is fair (has head
and tail) or fake (e.g. two heads). In the case of the coin
we would have to look at both sides to see if it is fair.
A QC can make a decision by looking only once (at the two
sides simultaneously).

\goodbreak
\topinsert
\centerline{\vbox{
\tenrm\baselineskip=10truept%
\noindent%
\KLEIN{Table 1.\ Input and output values of
{\eightsl constant} ($f_1(x)$, $f_2(x)$) and {\eightsl balanced} functions
($f_3(x)$, $f_4(x)$) of one input bit $x$.}
}}
\medskip
\centerline{\vbox{\offinterlineskip
\halign to\hsize{
\vrule#\tabskip=0pt plus1000pt minus 0pt&
\strut\hfil$#$\hfil&
\vrule#&
\hfil$#$\hfil&
\hfil$#$\hfil&
\vrule#&
\hfil$#$\hfil&
\hfil$#$\hfil
&\vrule#\tabskip=0pt \cr%
\noalign{\hrule}
height2pt&\omit&height2pt&\omit&\omit&height2pt&
\omit&\omit&height2pt\cr
\noalign{\hrule}
height4pt&\omit&height4pt&\omit&\omit&height4pt&
\omit&\omit&height4pt\cr
&x& &f_1(x)&f_2(x)& &f_3(x)&f_4(x)&\cr
height4pt&\omit&height4pt&\omit&\omit&height4pt&
\omit&\omit&height4pt\cr
\noalign{\hrule}
height2pt&\omit&height2pt&\omit&\omit&height2pt&
\omit&\omit&height2pt\cr
&0 &&  0  & 1 && 0 & 1 & \cr
&1 &&  0  & 1 && 1 & 0 & \cr
height2pt&\omit&height2pt&\omit&\omit&height2pt&
\omit&\omit&height2pt\cr
\noalign{\hrule}
height2pt&\omit&height2pt&\omit&\omit&height2pt&
\omit&\omit&height2pt\cr
\noalign{\hrule}
}}}
\endinsert

\goodbreak
\midinsert
\centerline{\vbox{
\tenrm\baselineskip=10truept%
\noindent%
\KLEIN{Table 2.\ Input and output values of
{\eightsl constant}
($f_1(x)$, $f_2(x)$) and
{\eightsl balanced}
functions
($f_3(x)$, $f_4(x)$, $f_5(x)$) of three input bits $x=\{x_1$, $x_2$, $x_3\}$.
Note that $f_4(x)$ only depends on $x_2$ and is therefore
rather trivially {\eightsl balanced}.}
}}
\medskip
\centerline{\vbox{\offinterlineskip
\halign to\hsize{
\vrule#\tabskip=0pt plus1000pt minus 0pt&
\strut\hfil$#$\hfil&
\hfil$#$\hfil&
\hfil$#$\hfil&
\vrule#&
\hfil$#$\hfil&
\hfil$#$\hfil&
\vrule#&
\hfil$#$\hfil&
\hfil$#$\hfil&
\hfil$#$\hfil
&\vrule#\tabskip=0pt \cr%
\noalign{\hrule}
height2pt&\omit&\omit&\omit&height2pt&\omit&\omit&height2pt&
\omit&\omit&\omit&height2pt\cr
\noalign{\hrule}
height4pt&\omit&\omit&\omit&height4pt&\omit&\omit&height4pt&
\omit&\omit&\omit&height4pt\cr
&x_1&x_2&x_3& &f_1(x_1,x_2,x_3)&f_2(x_1,x_2,x_3)& &
f_3(x_1,x_2,x_3)&f_4(x_1,x_2,x_3)&f_5(x_1,x_2,x_3)&\cr
height4pt&\omit&\omit&\omit&height4pt&\omit&\omit&height4pt&
\omit&\omit&\omit&height4pt\cr
\noalign{\hrule}
height2pt&\omit&\omit&\omit&height2pt&\omit&\omit&height2pt&
\omit&\omit&\omit&height2pt\cr
&0 & 0 & 0 &&  0  & 1 && 0 & 0 & 0& \cr
&1 & 0 & 0 &&  0  & 1 && 0 & 0 & 1& \cr
&0 & 1 & 0 &&  0  & 1 && 0 & 1 & 1& \cr
&1 & 1 & 0 &&  0  & 1 && 1 & 1 & 0& \cr
&0 & 0 & 1 &&  0  & 1 && 0 & 0 & 1& \cr
&1 & 0 & 1 &&  0  & 1 && 1 & 0 & 0& \cr
&0 & 1 & 1 &&  0  & 1 && 1 & 1 & 0& \cr
&1 & 1 & 1 &&  0  & 1 && 1 & 1 & 1& \cr
height2pt&\omit&\omit&\omit&height2pt&\omit&\omit&height2pt&
\omit&\omit&\omit&height2pt\cr
\noalign{\hrule}
height2pt&\omit&\omit&\omit&height2pt&\omit&\omit&height2pt&
\omit&\omit&\omit&height2pt\cr
\noalign{\hrule}
}}}
\endinsert

\goodbreak
\topinsert
\centerline{\vbox{
\tenrm\baselineskip=10truept%
\noindent%
\KLEIN{Table 3.\ Results of letting the sequences
$F_1,\ldots,F_3$ (see $(12)$) transform the four basis states.
Inspection of the outputs demonstrated that these
sequences implement the
{\eightsl constant} or {\eightsl balanced}
functions of Table 1.}
}}
\medskip
\centerline{\vbox{\offinterlineskip
\halign to\hsize{
\vrule#\tabskip=0pt plus1000pt minus 0pt&
\strut\hfil$#$\hfil&
\hfil$#$\hfil&
\vrule#&
\hfil$#$\hfil&
\hfil$#$\hfil&
\hfil$#$\hfil&
\hfil$#$\hfil&
&\vrule#\tabskip=0pt \cr%
\noalign{\hrule}
height2pt&\omit&\omit&height2pt&\omit&\omit&
\omit&\omit&height2pt\cr
\noalign{\hrule}
height4pt&\omit&\omit&height4pt&\omit&\omit&
\omit&\omit&height4pt\cr
&x&\KET{\Psi}&&F_1\KET{\Psi}&F_2\KET{\Psi}& F_3\KET{\Psi}&F_4\KET{\Psi}&\cr
height4pt&\omit&\omit&height4pt&\omit&\omit&
\omit&\omit&height4pt\cr
\noalign{\hrule}
height2pt&\omit&\omit&height2pt&\omit&\omit&
\omit&\omit&height2pt\cr
&0&\KET{00} &&  -\KET{00}  & i\KET{01} & \phantom{-}e^{-i\pi/4}\KET{00} & -e^{+i\pi/4}\KET{01} & \cr
&1&\KET{10} &&  -\KET{10}  & i\KET{11} &          - e^{-i\pi/4}\KET{11} & \phantom{-}e^{+i\pi/4}\KET{10} & \cr
&2&\KET{01} &&  -\KET{01}  & i\KET{00} & \phantom{-}e^{-i\pi/4}\KET{01} & -e^{+i\pi/4}\KET{00} & \cr
&3&\KET{11} &&  -\KET{11}  & i\KET{10}& -e^{-i\pi/4}\KET{10} & \phantom{-}e^{+i\pi/4}\KET{11} & \cr
height2pt&\omit&\omit&height2pt&\omit&\omit&
\omit&\omit&height2pt\cr
\noalign{\hrule}
height2pt&\omit&\omit&height2pt&\omit&\omit&
\omit&\omit&height2pt\cr
\noalign{\hrule}
}}}
\endinsert

\goodbreak
\topinsert
\centerline{\vbox{
\tenrm\baselineskip=10truept
\noindent%
\KLEIN{Table 4.\ Specification of the micro instructions
implementing the two-qubit NMR QC on the QCE.
Frequencies have been rescaled such that $h_{j,\alpha,0}=1$
corresponds to $500\hbox{MHz}$. The execution time of each
micro instruction is given by the second row ($\tau/2\pi$).
The inverse of e.g. ${\bar X}_1$ is found by reversing
the sign of $h_{1,x,1}$. Note that the QCE is constructed
such that a rotation about the $x$($y$) axis requires
a RF\-pulse along the $y$($x$) direction (see Appendix A).
}
}}
\medskip
\centerline{\vbox{\offinterlineskip
\halign to\hsize{
\vrule#\tabskip=0pt plus1000pt minus 0pt&
\strut\hfil$#$\hfil&
\vrule#&
\hfil$#$\hfil& 
\hfil$#$\hfil& 
\hfil$#$\hfil& 
\hfil$#$\hfil& 
\hfil$#$\hfil& 
\hfil$#$\hfil  
&\vrule#\tabskip=0pt \cr%
\noalign{\hrule}
height2pt&\omit&height2pt&\omit&\omit&\omit&\omit&\omit&\omit&height2pt\cr
\noalign{\hrule}
height4pt&\omit&height4pt&\omit&\omit&\omit&\omit&\omit&\omit&height4pt\cr
&\hbox{Parameter}&& X_1 & {\bar X}_2 & Y_1 & {\bar Y}_2 & I(\pi/2) &I(\pi)& \cr
height4pt&\omit&height4pt&\omit&\omit&\omit&\omit&\omit&\omit&height4pt\cr
\noalign{\hrule}
height2pt&\omit&height2pt&\omit&\omit&\omit&\omit&\omit&\omit&height2pt\cr
&\tau/2\pi &&  10  & 40    &  10   &  40   &  25\times10^{4} & 50\times10^{4} & \cr
height2pt&\omit&height2pt&\omit&\omit&\omit&\omit&\omit&\omit&height2pt\cr
\noalign{\hrule}
height2pt&\omit&height2pt&\omit&\omit&\omit&\omit&\omit&\omit&height2pt\cr
&J_{1,2,z}  &&-10^{-6}&-10^{-6}&-10^{-6}&-10^{-6}&-10^{-6}&-10^{-6}& \cr
height2pt&\omit&height2pt&\omit&\omit&\omit&\omit&\omit&\omit&height2pt\cr
\noalign{\hrule}
height2pt&\omit&height2pt&\omit&\omit&\omit&\omit&\omit&\omit&height2pt\cr
&h_{1,x,0}  &&  0  &  0  &  0  &  0  &   0       & 0    & \cr
&h_{2,x,0}  &&  0  &  0  &  0  &  0  &   0       & 0    & \cr
&h_{1,y,0}  &&  0  &  0  &  0  &  0  &   0       & 0    & \cr
&h_{2,y,0}  &&  0  &  0  &  0  &  0  &   0       & 0    & \cr
&h_{1,z,0}  &&  1  &  1  &  1  &  1  &   1       & 1    & \cr
&h_{2,z,0}  &&  0.25  &  0.25  &  0.25  &  0.25  &   0.25      & 0.25    & \cr
height2pt&\omit&height2pt&\omit&\omit&\omit&\omit&\omit&\omit&height2pt\cr
\noalign{\hrule}
height2pt&\omit&height2pt&\omit&\omit&\omit&\omit&\omit&\omit&height2pt\cr
&h_{1,x,1}  &&  0  &  0  &  0.05  &  -0.05  &   0       & 0    & \cr
&h_{2,x,1}  &&  0  &  0  &  0.0125  &  -0.0125  &   0       & 0    & \cr
&f_{1,x}  &&  0  &  0  &  1  &  0.25  &   0       & 0    & \cr
&f_{2,x}  &&  0  &  0  &  1  &  0.25  &   0       & 0    & \cr
&\varphi_{1,x}  &&  0  &  0  &  0  &  0  &   0       & 0    & \cr
&\varphi_{2,x}  &&  0  &  0  &  0  &  0  &   0       & 0    & \cr
height2pt&\omit&height2pt&\omit&\omit&\omit&\omit&\omit&\omit&height2pt\cr
\noalign{\hrule}
height2pt&\omit&height2pt&\omit&\omit&\omit&\omit&\omit&\omit&height2pt\cr
&h_{1,y,1}  && -0.05  & 0.05  &  0  &  0  &   0       & 0    & \cr
&h_{2,y,1}  && -0.0125  & 0.0125  &  0  &  0  &   0       & 0    & \cr
&f_{1,y}  &&  1  &  0.25  &  0  &  0  &   0       & 0    & \cr
&f_{2,y}  &&  1  &  0.25  &  0  &  0  &   0       & 0    & \cr
&\varphi_{1,y}  &&  0  &  0  &  0  &  0  &   0       & 0    & \cr
&\varphi_{2,y}  &&  0  &  0  &  0  &  0  &   0       & 0    & \cr
height2pt&\omit&height2pt&\omit&\omit&\omit&\omit&\omit&\omit&height2pt\cr
\noalign{\hrule}
height2pt&\omit&height2pt&\omit&\omit&\omit&\omit&\omit&\omit&height2pt\cr
&h_{1,z,1}  &&  0  &  0  &  0  &  0  &   0       & 0    & \cr
&h_{2,z,1}  &&  0  &  0  &  0  &  0  &   0       & 0    & \cr
&f_{1,z}  &&  0  &  0  &  0  &  0  &   0       & 0    & \cr
&f_{2,z}  &&  0  &  0  &  0  &  0  &   0       & 0    & \cr
&\varphi_{1,z}  &&  0  &  0  &  0  &  0  &   0       & 0    & \cr
&\varphi_{2,z}  &&  0  &  0  &  0  &  0  &   0       & 0    & \cr
height2pt&\omit&height2pt&\omit&\omit&\omit&\omit&\omit&\omit&height2pt\cr
\noalign{\hrule}
height2pt&\omit&height2pt&\omit&\omit&\omit&\omit&\omit&\omit&height2pt\cr
\noalign{\hrule}
}}}
\endinsert

\goodbreak
\topinsert
\centerline{\vbox{
\tenrm\baselineskip=10truept%
\noindent%
\KLEIN{Table 5.\ Final state of the QC after running
the D\-J algorithm for the case of the
ideal QC ($Q_1,Q_2$, see table 5)
and the NMR-QC ($\hat Q_1,\hat Q_2$, see table 4).
The results ($\tilde Q_1,\tilde Q_2$) have been obtained
by modifying the NMR MI's such that the
RF\-pulses only affect the spin that is in resonance.
The last two rows show the results of running
the refined version\REF{\COLLINSone} of the D\-J algorithm.
${\cal Q}_1$ : Ideal operations
${\hat{\cal Q}}_1$ : NMR implementation.
}
}}
\medskip
\centerline{\vbox{\offinterlineskip
\halign to\hsize{
\vrule#\tabskip=0pt plus1000pt minus 0pt&
\strut\hfil$#$\hfil&
\vrule#&
\hfil$#$\hfil&
\hfil$#$\hfil&
\hfil$#$\hfil&
\hfil$#$\hfil&
&\vrule#\tabskip=0pt \cr%
\noalign{\hrule}
height2pt&\omit&height2pt&\omit&\omit&\omit&\omit&height2pt\cr
\noalign{\hrule}
height4pt&\omit&height4pt&\omit&\omit&\omit&\omit&height4pt\cr
& &&f_1(x)&f_2(x)& f_3(x)&f_4(x)&\cr
height4pt&\omit&height4pt&\omit&\omit&\omit&\omit&height4pt\cr
\noalign{\hrule}
height2pt&\omit&height2pt&\omit&\omit&\omit&\omit&height2pt\cr
&Q_1     &&  0.000      & 0.000     & 1.000     & 1.000 & \cr
&Q_2     &&  0.000      & 0.000     & 0.000     & 0.000 & \cr
height2pt&\omit&height2pt&\omit&\omit&\omit&\omit&height2pt\cr
\noalign{\hrule}
height2pt&\omit&height2pt&\omit&\omit&\omit&\omit&height2pt\cr
&\hat Q_1&&  0.169  & 0.064 & 0.867 & 0.867 & \cr
&\hat Q_2&&  0.999  & 1.000 & 0.001 & 0.002 & \cr
height2pt&\omit&height2pt&\omit&\omit&\omit&\omit&height2pt\cr
\noalign{\hrule}
height2pt&\omit&height2pt&\omit&\omit&\omit&\omit&height2pt\cr
&\tilde Q_1&&  0.000  & 0.000 & 0.998 & 0.998 & \cr
&\tilde Q_2&&  1.000  & 1.000 & 0.001 & 0.001 & \cr
height2pt&\omit&height2pt&\omit&\omit&\omit&\omit&height2pt\cr
\noalign{\hrule}
height2pt&\omit&height2pt&\omit&\omit&\omit&\omit&height2pt\cr
&{\hat{\cal Q}}_1&&  0.000  & 0.000 & 1.000 & 1.000 & \cr
height2pt&\omit&height2pt&\omit&\omit&\omit&\omit&height2pt\cr
\noalign{\hrule}
height2pt&\omit&height2pt&\omit&\omit&\omit&\omit&height2pt\cr
&{\cal Q}_1     &&  0.000      & 0.000     & 0.995     & 0.996 & \cr
height2pt&\omit&height2pt&\omit&\omit&\omit&\omit&height2pt\cr
\noalign{\hrule}
height2pt&\omit&height2pt&\omit&\omit&\omit&\omit&height2pt\cr
\noalign{\hrule}
}}}
\endinsert

\goodbreak
\topinsert
\centerline{\vbox{
\tenrm\baselineskip=10truept
\noindent%
\KLEIN{Table 6.\ Specification of the micro instructions
implementing a mathematically perfect two-qubit QC on the QCE.
The execution time of each
micro instruction is given by the second row ($\tau/2\pi$).
The inverse of e.g. ${\bar X}_1$ is found by reversing
the sign of $h_{1,x,0}$. Model parameters omitted are
zero for all micro instructions.
}
}}
\medskip
\centerline{\vbox{\offinterlineskip
\halign to\hsize{
\vrule#\tabskip=0pt plus1000pt minus 0pt&
\strut\hfil$#$\hfil&
\vrule#&
\hfil$#$\hfil& 
\hfil$#$\hfil& 
\hfil$#$\hfil& 
\hfil$#$\hfil& 
\hfil$#$\hfil& 
\hfil$#$\hfil  
&\vrule#\tabskip=0pt \cr%
\noalign{\hrule}
height2pt&\omit&height2pt&\omit&\omit&\omit&\omit&\omit&\omit&height2pt\cr
\noalign{\hrule}
height4pt&\omit&height4pt&\omit&\omit&\omit&\omit&\omit&\omit&height4pt\cr
&\hbox{Parameter}&& X_1 & {\bar X}_2 & Y_1 & {\bar Y}_2 & I(\pi/2) &I(\pi)& \cr
height4pt&\omit&height4pt&\omit&\omit&\omit&\omit&\omit&\omit&height4pt\cr
\noalign{\hrule}
height2pt&\omit&height2pt&\omit&\omit&\omit&\omit&\omit&\omit&height2pt\cr
&\tau/2\pi &&  0.25  & 0.25   &  0.25   &  0.25   &  25\times10^{4} & 50\times10^{4} & \cr
height2pt&\omit&height2pt&\omit&\omit&\omit&\omit&\omit&\omit&height2pt\cr
\noalign{\hrule}
height2pt&\omit&height2pt&\omit&\omit&\omit&\omit&\omit&\omit&height2pt\cr
&J_{1,2,z}  &&  0  &  0  &  0  &  0  &   -10^{-6}& -10^{-6}& \cr
height2pt&\omit&height2pt&\omit&\omit&\omit&\omit&\omit&\omit&height2pt\cr
\noalign{\hrule}
height2pt&\omit&height2pt&\omit&\omit&\omit&\omit&\omit&\omit&height2pt\cr
&h_{1,x,0}  &&  +1  &  0  &  0  &  0  &   0       & 0    & \cr
&h_{2,x,0}  &&  0  &  -1  &  0  &  0  &   0       & 0    & \cr
&h_{1,y,0}  &&  0  &  0  &  +1  &  0  &   0       & 0    & \cr
&h_{2,y,0}  &&  0  &  0  &  0  &  -1  &   0       & 0    & \cr
height2pt&\omit&height2pt&\omit&\omit&\omit&\omit&\omit&\omit&height2pt\cr
\noalign{\hrule}
height2pt&\omit&height2pt&\omit&\omit&\omit&\omit&\omit&\omit&height2pt\cr
\noalign{\hrule}
}}}
\endinsert

In the NMR experiment the two qubits of the QC (i.e. the two
nuclear spins of the chloroform molecule) are used during the execution
of the D\-J algorithm although in principle only one qubit would do\REF{\COLLINSone}.
However our aim is to simulate the NMR\-QC experiment
and therefore we will closely follow Ref.\REF{\CHUANGtwo}.
Accordingly the first qubit is considered as the input variable, the other one
serves as work space.

Before the actual calculation starts the QC has to be initialized.
This amounts to setting each of the two qubits to \KET{0}.
On the QCE this is accomplished by the MI
``Initialize'', a reserved MI name in the QCE (see above).
The first step in the D\-J algorithm is to prepare the QC
by putting the first qubit in the state
$(\KET{0}+\KET{1})/\sqrt{2}$ and
the second one in
$(\KET{0}-\KET{1})/\sqrt{2}$\REF{\CHUANGtwo}.
This can be done by performing two rotations:

\FORM{\hbox{Prepare}\Leftrightarrow {\BAR Y}_2 {\NOBAR Y}_1},

where ${\NOBAR Y}_j$ represents the operation of rotating {\bf clock-wise}
the spin $j$ by $\pi/2$ along the $y$ axis,
and ${\BAR Y}_j$ its inverse (see Appendix A).
In this paper we adopt the convention that
all expressions like \CUR{0}/ have to be read from {\bf right to left}.

The next step is to compute the function $f(x)$.
Following\REF{\CHUANGtwo} the two {\sl constant} and two {\sl balanced} functions listed
in Table 1 can be implemented by the sequences

\EQA{
f_1(x) \Leftrightarrow &\, F_1={\NOBAR X}_2 {\NOBAR X}_2 I({\pi/2}) {\NOBAR X}_2 {\NOBAR X}_2 I({\pi/2})\CR,
f_2(x) \Leftrightarrow &\, F_2=I({\pi/2}){\NOBAR X}_2 {\NOBAR X}_2 I({\pi/2}) \CR,
f_3(x) \Leftrightarrow &\, F_3={\NOBAR Y}_1 {\BAR X}_1 {\BAR Y}_1 {\NOBAR X}_2 {\BAR Y}_2 I(\pi) {\NOBAR Y}_2 \CR,
f_4(x) \Leftrightarrow &\, F_4={\NOBAR Y}_1 {\BAR X}_1 {\BAR Y}_1 {\BAR X}_2 {\BAR Y}_2 I(\pi) {\NOBAR Y}_2
\CR,
}

where ${\NOBAR X}_j$ denotes
the {\bf clock-wise} rotation of spin $j$ by $\pi/2$ along the $x$ axis,
${\BAR X}_j$ the inverse operation and
$I(a)\equiv e^{-ia S_1^zS_2^z}$
represents the time evolution due to $H$ itself.
In Table 3 we show the result of letting the sequences \CUR{0}/ act
on the basis states. It is clear that they have the desired properties.
Note that prefactors have no physical relevance (they drop out when
we compute expectation values) and that
$F_1$ is a rather complicated version of the identity operation.

Finally there is a read\-out operation which corresponds to
in the inverse of the ``Prepare'':

\FORM{\hbox{ReadOut}\Leftrightarrow {\BAR Y}_1 {\NOBAR Y}_2}.

Note that there is some flexibility in the choice of these
sequences. For instance to ``Prepare'' we could
have used Walsh\-Hadamard (WH) transformations ${\NOBAR W}_1 {\NOBAR W}_2$
as well.

Upto this point the D\-J algorithm has been written as a sequence
of unitary operations that perform specific tasks.
Now we consider two different implementations of these
unitary transformations: The first one will be physical, i.e.
we will use the QCE simulate the NMR\-QC experiment itself.
The second will be ``computer\-science'' like, i.e. we will
use highly idealized, non\-realizable rotations.

NMR uses radiofrequency electromagnetic
pulses to rotate the spins\REF{
\REFD{C.P. Slichter,
``Principles of Magnetic Resonance'',
(Springer, Berlin, 1990)/}{\SLICHTER},
\REFD{G. Baym, ``Lectures on Quantum Mechanics'',
(W.A. Bejamin, Reading MA, 1974)/}{\BAYM}
}.
By tuning the frequency
of the RF-field to the precession frequency of a particular
spin, the power of the applied pulse (= intensity times duration)
controls how much the spin will rotate. The axis of the
rotation is determined by the direction of the applied RF-field
(see\REF{\SLICHTER,\BAYM} or Appendix A).
A possible choice of the model parameters, corresponding
to the actual experimental values of these parameters,
is given in Table 4. For simplicity all frequencies
have been normalized with respect to the largest
one (i.e. $500\hbox{MHz}$ in the experiments \REF{\CHUANGtwo,\CHUANGthree}) .
Also note that it is convenient to express execution times
in units of $2\pi$, the default setting in the QCE.

The results of running the QCE with the MI's
simulating the NMR experiment are summarized in Fig.1.
The first qubit ($Q_j=1/2-\EXPECT{S_j^z}$)
unambigously tells us that the functions $f_1(x)$ and $f_2(x)$ are constant and that
$f_3(x)$ and $f_4(x)$ are balanced.
Clearly the QCE qualitatively reproduces the experimental results.
In the D\-J algorithm the final state of the second qubit is irrelevant.
In the final state the numerical value of qubit 1,
is only approximately zero or one (see Table 5). This is a direct consequence
of the fact that we are simulating a genuine physical system.

In an NMR experiment, application of a RF\-pulse affects all spins in the sample.
Although the response of a spin to the RF-field
will only be large when this spin is at resonance, the state of the spins that are not
in resonance will also change. These unitary transformations
not necessarily commute with the sequence of unitary transformations that
follow and may therefore affect the final outcome of the whole computation.
Furthermore the use of a time-dependent external field to rotate spins
is only an approximation to the simple rotations envisaged in theoretical work
(see Appendix A). This definitely has an effect on the expectation
values of the spin operators.

With the QCE it is very easy to make
a detailed comparison between physical and idealized
implementations of QC's: We simply replace the set of MI's (``NMR'')
by another one (``Ideal'', or ``NMR-Ideal'') and re\-run the QP's
by simply clicking on the execute buttons.
The model parameters we have chosen to implement the ``ideal'' operations
are listed in Table 6.
The set ``NMR-Ideal'' is a copy of ``NMR'' (see Table 4) except
that the RF\-pulses only affect the spin that is in resonance,
i.e. all operations on qubit $j$ have $h_{2,x,j}=h_{2,y,j}=0$.

The results of executing the QA's for the ``Ideal'' case are shown in Fig.2.
In the final state the qubits are exactly \KET{0} or \KET{1}, as expected.
The state of the second qubit not always matches the corresponding state
of Fig.1. As mentioned above this is due to the approximate nature
of the operations used in the NMR case, but
as the final state of the second qubit is irrelevant for the D\-J algorithm
there is no problem.
In Table 5 we collect the numerical
values of the qubits as obtained by running
the D\-J algorithm on the NMR and ideal QC.
It is clear in that all cases the D\-J algorithm gives the correct answer.

\SECTION{Collins-Kim-Holton algorithm}

>From the description of the DJ algorithm of one variable
(\REF{\CHUANGtwo}, Fig.1) it is evident that the second qubit is redundant
because the function call (step T2 in \REF{\CHUANGtwo})
leaves the state of the second qubit, i.e. the work space, untouched.
A refined version of the D\-J algorithm (for an arbitrary number of qubits)
that does not require a bit for the evaluation of the function
is given in\REF{\COLLINSone}.
For one variable, the QC has to compute the function

\FORM{f_j \Leftrightarrow \sum_{x=0}^1 (-1)^{f_j(x)}\KET{x}}.

Following\REF{\COLLINSone} this may be accomplished via
an $f$\-controlled gate defined by

\FORM{U_f \KET{x}= (-1)^{f(x)}\KET{x}}.

Accordingly, once choice (there are several) of the set of sequences that implements the
refined version of the D\-J algorithm reads:

\EQA{
\hbox{Prepare}\Leftrightarrow &\,{\BAR Y}_1\CR,
f_1(x) \Leftrightarrow &\, F_1={\NOBAR W}_1 {\NOBAR W}_1\CR,
f_2(x) \Leftrightarrow &\, F_2={\NOBAR X}_1 {\NOBAR X}_1 \CR,
f_3(x) \Leftrightarrow &\, F_3={\BAR Y}_1 {\NOBAR X}_1 {\NOBAR Y}_1 {\BAR Y}_1 {\NOBAR X}_1 {\NOBAR Y}_1 \CR,
f_4(x) \Leftrightarrow &\, F_4={\NOBAR Y}_1 {\BAR X}_1 {\BAR Y}_1 {\NOBAR Y}_1 {\BAR X}_1 {\BAR Y}_1 \CR,
\hbox{ReadOut}\Leftrightarrow &\,{\BAR Y}_1\CR,
}

The results of running these QP's on the QCE are given in Table 6.
It is clear that the refined version performs as expected.

\SECTION{Grover's database search algorithm}

On a conventional computer finding a particular entry in an unsorted
list of $N$ elements requires of the order of $N$ operations.
Grover has shown that a QC can find the item using
only \ORDER{\sqrt{N}} attempts\REF{\GROVERzero,\GROVERone}.
Consider the extremely simple case of a database containing
four items and functions $g_j(x)$, $j=0,\ldots,3$ that upon query of the database
return minus one if $x=j$ and plus one if $x\not=j$.
Assuming a uniform probability distribution for the item
to be in one of the four locations, the average number
of queries required by a conventional algorithm is 9/4.
With Grover's QA the correct answer can be found
in a single query (this result only holds for
a database with 4 items).
Grover's algorithm for the four\-item database can be
implemented on a two\-qubit QC.

The key ingredient of Grover's algorithm is an
operation that replaces each amplitude
of the basis states in the superposition by
two times the average amplitude minus
the amplitude itself.
This operation is called ``inversion about the mean'' and
amplifies the amplitude of the basis state that represents the
searched\-for item.
To see how this works it is useful to consider an example.
Let us assume that the item to search for corresponds
to e.g. number 2 ($g_2(0)=g_2(1)=g_2(3)=1$ and $g_2(2)=-1$).
Using the binary representation of integers with the order of the bits
reversed,
the QC is in the state (up to an irrelevant phase factor as usual)

\FORM{\KET{\Psi}={1\over 2}(\KET{00}+\KET{10}-\KET{01}+\KET{11})
}.

We return to the question of how to prepare this state below.
The operator $D$ that inverts states like \CUR{0}/ about their mean reads

\FORM{D={1\over2}\pmatrix{
-1&\phantom{-}1&\phantom{-}1&\phantom{-}1\cr
\phantom{-}1&-1&\phantom{-}1&\phantom{-}1\cr
\phantom{-}1&\phantom{-}1&-1&\phantom{-}1\cr
\phantom{-}1&\phantom{-}1&\phantom{-}1&-1\cr}
\quad;\quad\matrix{\KET{00}\cr\KET{10}\cr\KET{01}\cr\KET{11}\cr}
}.

The mean amplitude of \CUR{-1}/ is 1/4 and we find that

\EQA{
D\KET{\Psi}=&\KET{01}\CR,
\NOALIGN{i.e. the correct answer, and}
D^2\KET{\Psi}=&
{1\over 2}(\KET{00}+\KET{10}+\KET{01}+\KET{11})\CR,
\NOALIGN{}
D^3\KET{\Psi}=&
-{1\over 2}(\KET{00}+\KET{10}-\KET{01}+\KET{11})=-\KET{\Psi}\CR,
}

showing that (in the case of 2 qubits)
the correct answer (i.e. the absolute value of the amplitude of \KET{10} equal
to one) is obtained after 1, 4, 7, ... iterations.
In general, for more than two qubits,
more than one application of $D$ is required to get the correct answer.
In this sense the 2\-qubit case is somewhat special.

The next task is to express the preparation and query steps
in terms of elementary rotations. For illustrative purposes
we stick to the example used above.
Initially we set the QC in the state \KET{00}, i.e.
the state with both spins up\REF{
\REFD{We follow the convention used earlier in this paper,
i.e. the one used in\REF{\CHUANGtwo} and
therefore deviate from the notation used in\REF{\CHUANGthree}/}{\XXXXXX}}.
and then transform \KET{00} to the linear superposition \CUR{-2}/
by a two\-step process.
First we set the QC in the uniform superposition state \KET{U}:

\FORM{
\hbox{Prepare}\Leftrightarrow \KET{U}\equiv W_2W_1\KET{00}
=-{1\over 2}(\KET{00}+\KET{10}+\KET{01}+\KET{11})
},

where

\FORM{W_j=X_jX_j{\bar Y}_j=
-{\bar X}_j {\bar X}_j{\bar Y}_j=
{i\over\sqrt{2}}\pmatrix{1&\phantom{-}1\cr1&-1\cr}_j
},

is the WH tranform on qubit $j$ which transforms
\KET{0} to $i(\KET{0}+\KET{1})/\sqrt{2}$
(see Appendix A).
The transformation that corresponds to
the application of $g_2(x)$ to the uniform superposition state is

\FORM{F_2=\pmatrix{
\phantom{-}1&\phantom{-}0&\phantom{-}0&\phantom{-}0\cr
\phantom{-}0&\phantom{-}1&\phantom{-}0&\phantom{-}0\cr
\phantom{-}0&\phantom{-}0&-1&\phantom{-}0\cr
\phantom{-}0&\phantom{-}0&\phantom{-}0&\phantom{-}1\cr}
\quad;\quad\matrix{\KET{00}\cr\KET{10}\cr\KET{01}\cr\KET{11}\cr}
}.

This transformation can be implemented by first letting the
system evolve in time:

\EQA{
I(\pi)\KET{U}= e^{-i\pi S_1^z S_2^z}
&\left[
{1\over 2}(\KET{00}+\KET{10}+\KET{01}+\KET{11})
\right]\cr
&=
{1\over 2}(e^{-i\pi/4}\KET{00}+e^{+i\pi/4}\KET{10}
+e^{+i\pi/4}\KET{01}+e^{-i\pi/4}\KET{11})\CRZ.
}

For the NMR-QC based on hamiltonian \HAM\ this means
letting the system evolve in time (without applying pulses)
for a time $\tau_0 =-\pi/J_{1,2,z}$ (recall $J_{1,2,z}<0$).
Next we apply a sequence of single\-spin rotations
to change the four phase factors such that we get the desired
state. The two sequences
$Y X {\bar Y}$ and $Y{\bar X}{\bar Y}$ (see Appendix B)
are particulary useful for this purpose.
We find

\EQA{
Y_1 X_1 {\bar Y}_1 Y_2{\bar X}_2{\bar Y}_2
&\left[
{1\over 2}(e^{-i\pi/4}\KET{00}+e^{+i\pi/4}\KET{10}
+e^{+i\pi/4}\KET{01}+e^{-i\pi/4}\KET{11})
\right]\cr
&=
{1\over 2}(e^{-i\pi/4}\KET{00}+e^{-i\pi/4}\KET{10}
+e^{+3i\pi/4}\KET{01}+e^{-i\pi/4}\KET{11})\cr
&=
{e^{-i\pi/4}\over 2}(\KET{00}+\KET{10}
-\KET{01}+\KET{11})\CRZ.
}

Combining \CUR{-1}/ and \CUR{0}/ we can construct the sequence $G_j$ that
transforms the uniform superposition \KET{U}
to the state that corresponds to $g_j(x)$:

\EQA{
F_0=&Y_1 {\bar X_1} {\bar Y}_1 Y_2 {\bar X_2}{\bar Y}_2 I(\pi)\CR,
F_1=&Y_1 {\bar X_1} {\bar Y}_1 Y_2 {     X_2}{\bar Y}_2 I(\pi)\CR,
F_2=&Y_1 {     X_1} {\bar Y}_1 Y_2 {\bar X_2}{\bar Y}_2 I(\pi)\CR,
F_3=&Y_1 {     X_1} {\bar Y}_1 Y_2 {     X_2}{\bar Y}_2 I(\pi)\CR.
}

The remaining task is to express the operation of inversion about the mean,
i.e. the matrix $D$ (see \CUR{-7}/), by a sequence of
elementary operations. It is not difficult to see that
$D$ can be written as the product of a
WH transform, a conditional phase shift $P$ and another WH transform:

\EQA{D=&W_1 W_2 P W_1 W_2\cr
=&W_1 W_2
\pmatrix{
\phantom{-}1&\phantom{-}0&\phantom{-}0&\phantom{-}0\cr
\phantom{-}0&-1&\phantom{-}0&\phantom{-}0\cr
\phantom{-}0&\phantom{-}0&-1&\phantom{-}0\cr
\phantom{-}0&\phantom{-}0&\phantom{-}0&-1\cr}
W_1 W_2 \CR.}

The same approach that was used to implement $g_2(x)$ also works
for the conditional phase shift $P$ ($=-F_0$)
and yields

\FORM{P=
Y_1{\bar X}_1{\bar Y}_1 Y_2 {\bar X}_2{\bar Y}_2 I(\pi)}.

The complete sequence $U_j$ reads

\FORM{
U_j=W_1 W_2 P W_1 W_2 F_j
}.

Each sequence $U_j$ can be shortened by
observing that in some cases a rotation is followed by its inverse.
Making use of the alternative representations
of the WH transform $W_i$ (see (Appendix B)), the sequence for e.g. $j=1$
can be written as

\EQA{
W_1 W_2 F_1=&
-{     X}_1 {     X}_1{\bar Y}_1 {\bar X}_2 {\bar X}_2{\bar Y}_2
Y_1 {\bar X}_1 {\bar Y}_1 Y_2{     X}_2{\bar Y}_2 I(\pi) \cr
=&
-{ X}_1 {\bar Y}_1 {\bar X}_2 {\bar Y}_2 I(\pi)    \CRZ.
}

The sequences for the other cases can be shortened as well,
yielding

\EQA{
U_0=&{     X}_1{\bar Y}_1 {     X}_2{\bar Y}_2 I(\pi)
{     X}_1 {\bar Y}_1 {     X}_2 {\bar Y}_2 I(\pi)\CR,
U_1=&{     X}_1{\bar Y}_1 {     X}_2{\bar Y}_2 I(\pi)
{     X}_1 {\bar Y}_1 {\bar X}_2 {\bar Y}_2 I(\pi)\CR,
U_2=&{     X}_1{\bar Y}_1 {     X}_2{\bar Y}_2 I(\pi)
{\bar X}_1 {\bar Y}_1 {     X}_2 {\bar Y}_2 I(\pi)\CR,
U_3=&{     X}_1{\bar Y}_1 {     X}_2{\bar Y}_2 I(\pi)
{\bar X}_1 {\bar Y}_1 {\bar X}_2 {\bar Y}_2 I(\pi)\CR,
}

where in $U_1$ and $U_2$ a physically irrelevant
sign has been dropped.
Note that the binary representation of $x$ translates into the presence (0) or absence
(1) in \CUR{0}/ of a bar on the rightmost $X_1$ and $X_2$.

As before, our aim is to use the QCE to simulate the NMR\-QC
experiment\REF{\CHUANGthree}.
For the D\-J algorithm we already specified
the physical parameters for the elementary operations
and we will make use of the same set of MI's here.
In Figs.3 and 4 we show the QCE after running the four cases $g_0(x),\ldots,g_3(x)$
using the NMR (Fig.4) MI's.
The numerical values of the qubits in the final state are given in Table 7,
for the ideal and NMR QC.
In both cases the QA performs as it should. In the ideal case, the
final state of the QC is exactly equal to \KET{x} (binary representation
of integers). Using RF\-pulses instead of ideal transformations
to perform $\pi/2$ rotations leads to less certain answers: The final
state is no longer a pure basis state but some linear superposition of the four
basis states. What is beyond doubt though is that in all cases
the weight of \KET{x} is by far the largest.
Hence the QC returns the correct answer.

\goodbreak
\topinsert
\centerline{\vbox{
\tenrm\baselineskip=10truept%
\noindent%
\KLEIN{Table 7.\ Final state of the QC after running
the Grover's database search algorithm for the case of the
ideal QC ($Q_1,Q_2$, see table 5)
and the NMR-QC ($\hat Q_1,\hat Q_2$, see table 4).}
}}
\medskip
\centerline{\vbox{\offinterlineskip
\halign to\hsize{
\vrule#\tabskip=0pt plus1000pt minus 0pt&
\strut\hfil$#$\hfil&
\vrule#&
\hfil$#$\hfil&
\hfil$#$\hfil&
\hfil$#$\hfil&
\hfil$#$\hfil&
&\vrule#\tabskip=0pt \cr%
\noalign{\hrule}
height2pt&\omit&height2pt&\omit&\omit&
\omit&\omit&height2pt\cr
\noalign{\hrule}
height4pt&\omit&height4pt&\omit&\omit&
\omit&\omit&height4pt\cr
& &&g_0(x)&g_1(x)& g_2(x)&g_3(x)&\cr
height4pt&\omit&height4pt&\omit&\omit&
\omit&\omit&height4pt\cr
\noalign{\hrule}
height2pt&\omit&height2pt&\omit&\omit&
\omit&\omit&height2pt\cr
&Q_1     &&  0      & 1     & 0     & 1 & \cr
&Q_2     &&  0      & 0     & 1     & 1 & \cr
height2pt&\omit&height2pt&\omit&\omit&
\omit&\omit&height2pt\cr
\noalign{\hrule}
height2pt&\omit&height2pt&\omit&\omit&
\omit&\omit&height2pt\cr
&\hat Q_1&&  0.028  & 0.966 & 0.037 & 0.955 & \cr
&\hat Q_2&&  0.163  & 0.171 & 0.836 & 0.830 & \cr
height2pt&\omit&height2pt&\omit&\omit&
\omit&\omit&height2pt\cr
\noalign{\hrule}
height2pt&\omit&height2pt&\omit&\omit&
\omit&\omit&height2pt\cr
\noalign{\hrule}
}}}
\endinsert

\CHAPTER{Summary}

We have described the internal operation of QCE, a the software tool
for simulating hardware realizations of quantum computers.
The QCE simulates the physical (quantum) processes
that govern the operation of the hardware quantum processor
by solving the time\-dependent Schr\"odinger equation.
The use of the QCE has been illustrated by several
implementations of the Deutsch\-Jozsa
and Grover's database search algorithm,
on QC's using ideal and more realistic units,
such as those of 2-qubit NMR\-QC's.
Currently the QCE is used to study the stability of quantum
computers in relation to the non\-idealness of realizable
elementary operations\REF{
\REFD{H. De Raedt, A.H. Hams, K. Michielsen, S. Miyashita, and K. Saito,
``Quantum Spins Dynamics and Quantum Computation'',
J. Phys. Soc. Jpn. (in press), e-print {quant-ph/9911038}/}{\QCtwo}
}.

\CHAPTER{Acknowledgement}

Generous support from the Dutch ``Stichting Nationale Computer
Faciliteiten (NCF)''
is gratefully acknowledged.

\NOCHAPFORMNUMBERfalse 
\APPENDIX{Spin-1/2 algebra}

Here we present a collection of standard results on spin-1/2 systems
which are used in the paper and are taken from\REF{\BAYM}.
We begin with some notation.

The two basis states spanning the Hilbert space of a two-state quantum system
are usually denoted by

\FORM{\KET{\uparrow}\equiv\pmatrix{1\cr0}\quad,\quad
\KET{\downarrow}\equiv\pmatrix{0\cr1}
}.

The three components of the spin-1/2 operator ${\vec S}$
acting on this Hilbert space are defined by

\FORM{
S^x={\hbar\over2}\pmatrix{0&1\cr 1&0\cr}\quad,\quad
S^y={\hbar\over2}\pmatrix{0&-i\cr i&\phantom{-}0\cr}\quad,\quad
S^z={\hbar\over2}\pmatrix{1&\phantom{-}0\cr 0&-1\cr}
}.

By convention the represenation \CUR{0}/ is chosen such that
\KET{\uparrow} and \KET{\downarrow}
are eigenstates of $S^z$ with eigenvalues $+\hbar/2$ and $-\hbar/2$
respectively.

>From \CUR{0}/ it is clear that
$\left( S^x \right)^2=\left( S^y \right)^2=\left( S^z \right)^2=\hbar^2/4$
so that

\EQA{
\cos({\varphi S^\alpha\over\hbar})=&\cos({\varphi\over2})\pmatrix{1&0\cr0&1\cr}\CR,
\NOALIGN{and}
\sin({\varphi S^\alpha\over\hbar})=&{2\over\hbar}\sin({\varphi\over2})\,S^\alpha\CR.
}

The commutation relations between the three spin-components read

\FORM{\left[ S^\alpha,S^\beta \right] = i\hbar \epsilon_{\alpha\beta\gamma} S^\gamma
},

where $[A,B]\equiv AB-BA$,
$\epsilon_{\alpha\beta\gamma}$ is the totally antisymmetric unit tensor (
$\epsilon_{xyz}=\epsilon_{yzx}=\epsilon_{zxy}=1$,
$\epsilon_{\alpha\beta\gamma}=-\epsilon_{\beta\alpha\gamma}
=-\epsilon_{\gamma\beta\alpha}=-\epsilon_{\alpha\gamma\beta}$,
$\epsilon_{\alpha\alpha\gamma}=0$)
and the summation convention is assumed.

Rotation of the spin about an angle $\varphi$ around the axis $\beta$
gives

\FORM{S^\alpha(\varphi,\beta)\equiv
e^{i\varphi S^\beta/\hbar} S^\alpha e^{-i\varphi S^\beta/\hbar} =
S^\alpha \cos\varphi+\epsilon_{\alpha\beta\gamma} S^\gamma \sin\varphi
}.

Of particular interest to quantum computing are rotations about $\pi/2$ around the
$x$ and $y$\-axis defined by

\EQA{{\NOBAR X}\equiv& e^{i\pi S^x/2\hbar}={1\over\sqrt{2}}\pmatrix{1&i\cr i&1\cr}\CR,
\NOALIGN{and}
{\NOBAR Y}\equiv& e^{i\pi S^y/2\hbar}={1\over\sqrt{2}}\pmatrix{\phantom{-}1&1\cr -1&1\cr}\CR.}

The inverse of a rotation $Z$ will be denoted as ${\bar Z}$ and
if more than one spin is involved a subscript will be attached.
With our convention $\BRA{\uparrow} {\BAR Y}S^x{\NOBAR Y}
\KET{\uparrow}=-1/2$
so that a positive angle corresponds to a rotation in the clock\-wise direction.

Another basic operation is the Walsh-Hadamard transform $W$ which
rotates the state \KET{\uparrow} into
$(\KET{\downarrow} + \KET{\uparrow})/\sqrt{2}$
(up to an irrelevant phase factor), i.e. the uniform superposition
state.
In terms of elementary rotations the Walsh-Hadamard transform
reads

\FORM{W=X^2\bar Y=Y X^2
=-\bar X^2\bar Y=-Y \bar X^2 ={i\over\sqrt{2}}
\pmatrix{1&\phantom{-}1\cr 1&-1\cr}
},

For example

\FORM{W\KET{\uparrow}=
W\pmatrix{1\cr0\cr}
={i\over\sqrt{2}}\pmatrix{1\cr1\cr}
}.

We now consider the time evolution of a single spin subject to
a constant magnetic field along the $z$-axis and a RF-field
along the $x$-axis, i.e. the elementary model of NMR.
The TDSE reads

\FORM{i\hbar{\partial\over\partial t}\KET{\Phi(t)}=
-\left[ H_0 S^z + H_1 S^x \sin \omega t \right] \KET{\Phi(t)}
},

where $\KET{\Phi(t=0)}$ is the initial state of the two-state system
and we have set the phase in \FULLHAM\ to zero for notational convenience.
Substituting $\KET{\Phi(t)}=e^{it\omega_0 S^z/\hbar}\KET{\Psi(t)}$ yields

\FORM{i\hbar{\partial\over\partial t}\KET{\Psi(t)}=
-\left[ (H_0-\omega_0) S^z
+ H_1 S^x \sin\omega t\cos\omega_0 t
+ H_1 S^y \sin\omega t\sin\omega_0 t
\right] \KET{\Psi(t)}
},

which upon chosing $\omega_0=H_0$ can be written as

\FORM{i\hbar{\partial\over\partial t}\KET{\Psi(t)}=
-H_1\left[ S^x \sin\omega t\cos H_0 t
+ S^y \sin\omega t\sin H_0 t
\right] \KET{\Psi(t)}
}.

At resonance, i.e. $\omega=H_0$, we find

\FORM{i\hbar{\partial\over\partial t}\KET{\Psi(t)}=
-{H_1\over2}\left[ S^y
+ S^x \sin 2H_0 t
- S^y \cos 2H_0 t
\right] \KET{\Psi(t)}
}.

Assuming that the effects of the higher harmonic terms
(i.e. the terms in $\sin 2H_0 t$ and $\cos 2H_0 t$)
are small\REF{\BAYM} we obtain

\FORM{i\hbar{\partial\over\partial t}\KET{\Psi(t)}\approx
-{H_1\over2}S^y \KET{\Psi(t)}
},

which is easily solved to give

\FORM{
\KET{\Psi(t)}\approx e^{i t H_1 S^y/2\hbar}
\KET{\Psi(t=0)}
},

so that the overall action of an RF-pulse of duration $\tau$
can be written as

\FORM{
\KET{\Phi(t+\tau)}\approx e^{i\tau H_0 S^z/\hbar} e^{i \tau H_1 S^y/2\hbar}
\KET{\Phi(t)}
}.

>From \CUR{0}/ it follows that application of an RF-pulse of ``power''
$\tau H_1=\pi$ will have the effect of rotating the spin by
an angle of $\pi/2$ about the $y$-axis. For example

\EQA{e^{i\tau H_0 S^z/\hbar} e^{i \pi S^y/2\hbar}\KET{\uparrow}
=&e^{i\tau H_0 S^z/\hbar}\left[
\cos{\pi\over4}\pmatrix{1&0\cr 0&1\cr}+
i\sin{\pi\over4}\pmatrix{0&-i\cr i&\phantom{-}0\cr}
\right]\pmatrix{1\cr0\cr} \cr
=&{1\over\sqrt{2}}e^{i\tau H_0 S^z/\hbar}\pmatrix{\phantom{-}1\cr-1\cr} \cr
=&{1\over\sqrt{2}}\pmatrix{\phantom{-}e^{i\tau H_0/2}\cr -e^{-i\tau H_0 /2}\cr} \CRZ.}

In this rotated state the expectation values of the spin components are
given by

\EQA{
\BRACKET{\uparrow}{ e^{-i \pi S^y/2\hbar} e^{-i\tau H_0 S^z/\hbar} S^x
e^{i\tau H_0 S^z/\hbar} e^{i \pi S^y/2\hbar} |\uparrow} = & -\hbar\cos\tau H_0 \CR,
\NOALIGN{}
\BRACKET{\uparrow}{ e^{-i \pi S^y/2\hbar} e^{-i\tau H_0 S^z/\hbar} S^y
e^{i\tau H_0 S^z/\hbar} e^{i \pi S^y/2\hbar} |\uparrow} = & -\hbar\sin\tau H_0 \CR,
\NOALIGN{}
\BRACKET{\uparrow}{ e^{-i \pi S^y/2\hbar} e^{-i\tau H_0 S^z/\hbar} S^z
e^{i\tau H_0 S^z/\hbar} e^{i \pi S^y/2\hbar} |\uparrow} = & 0  \CR,
}

showing that the time of the RF\-pulse also affects the projection of the spin
on the $x$ and $y$ axis.

It is instructive to derive the TDSE that
corresponds to approximation \CUR{-2}/.
Taking the derivative of \CUR{-2}/ with respect to $t$ we obtain

\FORM{i\hbar{\partial\over\partial t}\KET{\Phi(t)}=
-H_0 S^z
-{H_1\over2} \left[ S^x \sin H_0 t + S^y \cos H_0 t\right]
\KET{\Phi(t)}
},

telling us that the approximate solution \CUR{-3}/ is the
exact solution for an RF field rotating in space\REF{\BAYM}.
The fact that the application of an RF-pulse does not
{\sl exactly} correspond to a simple rotation in spin space
may well be important for applications of NMR techniques to QC's.

Finally we note that our choice of using a ``$\sin\omega t$''
instead of ``$\cos\omega t$''\REF{\BAYM} to couple
the spin to the RF-field merely leads to a phase shift.
In the former case rotating the spin around the $x$-axis requires
a pulse along the $y$\-axis, whereas in the latter the pulse
should be applied along the $x$\-axis\REF{\BAYM}.

\APPENDIX{Basic operations}

Below we list a number of identities that are useful to
compute by hand the action of the sequences appearing above.
The convention adopted in this paper is that

\FORM{\KET{0}=\KET{\uparrow}=\pmatrix{1\cr0}\quad;\quad
\KET{1}=\KET{\downarrow}=\pmatrix{0\cr1}}.

A straightforward calculation yields:

\EQA{
{\NOBAR X}\KET{0}=&{1\over\sqrt{2}}(\KET{0}+i\KET{1}) \quad,\quad
{\NOBAR X}\KET{1}={1\over\sqrt{2}}(i\KET{0}+\KET{1}) \CR,
{\BAR X}\KET{0}=&{1\over\sqrt{2}}(\KET{0}-i\KET{1}) \quad,\quad
{\BAR X}\KET{1}={1\over\sqrt{2}}(-i\KET{0}+\KET{1}) \CR,
}

\EQA{
{\NOBAR Y}\KET{0}=&{1\over\sqrt{2}}(\KET{0}-\KET{1}) \quad,\quad
{\NOBAR Y}\KET{1}={1\over\sqrt{2}}(\KET{0}+\KET{1}) \CR,
{\BAR Y}\KET{0}=&{1\over\sqrt{2}}(\KET{0}+\KET{1}) \quad,\quad
{\BAR Y}\KET{1}={1\over\sqrt{2}}(-\KET{0}+\KET{1}) \CR,
}

\EQA{
{\NOBAR Y}\left[{1\over\sqrt{2}}(\KET{0}-\KET{1})\right] =& -\KET{1}\quad,\quad
{\NOBAR Y}\left[{1\over\sqrt{2}}(\KET{0}+\KET{1})\right] = \KET{0}\CR,
{\BAR Y}\left[{1\over\sqrt{2}}(\KET{0}-\KET{1})\right] =& \KET{0}\quad,\quad
{\BAR Y}\left[{1\over\sqrt{2}}(\KET{0}+\KET{1})\right] = \KET{1}\CR,
}

\EQA{
{\NOBAR X}{\NOBAR X}\KET{0}=&i\KET{1} \quad,\quad
{\NOBAR X}{\NOBAR X}\KET{1}=i\KET{0} \CR,
{\NOBAR Y}{\NOBAR Y}\KET{0}=&-\KET{1} \quad,\quad
{\NOBAR Y}{\NOBAR Y}\KET{1}=\KET{0} \CR,
}

\EQA{
{\NOBAR X}{\NOBAR Y}\KET{0}=&{e^{-i\pi/4}\over\sqrt{2}}(\KET{0}-\KET{1}) \quad,\quad
{\NOBAR X}{\NOBAR Y}\KET{1}={e^{+i\pi/4}\over\sqrt{2}}(\KET{0}+\KET{1}) \CR,
{\BAR X}{\NOBAR Y}\KET{0}=&{e^{+i\pi/4}\over\sqrt{2}}(\KET{0}-\KET{1}) \quad,\quad
{\BAR X}{\NOBAR Y}\KET{1}={e^{-i\pi/4}\over\sqrt{2}}(\KET{0}+\KET{1}) \CR,
{\NOBAR X}{\BAR Y}\KET{0}=&{e^{+i\pi/4}\over\sqrt{2}}(\KET{0}+\KET{1}) \quad,\quad
{\NOBAR X}{\BAR Y}\KET{1}={e^{-i\pi/4}\over\sqrt{2}}(-\KET{0}+\KET{1}) \CR,
{\BAR X}{\BAR Y}\KET{0}=&{e^{-i\pi/4}\over\sqrt{2}}(\KET{0}+\KET{1}) \quad,\quad
{\BAR X}{\BAR Y}\KET{1}={e^{+i\pi/4}\over\sqrt{2}}(-\KET{0}+\KET{1}) \CR,
}

\EQA{
{\NOBAR Y}{\NOBAR X}{\BAR Y}\KET{0}=&e^{+i\pi/4}\KET{0} \quad,\quad
{\NOBAR Y}{\NOBAR X}{\BAR Y}\KET{1}=e^{-i\pi/4}\KET{1} \CR,
{\BAR Y}{\NOBAR X}{\NOBAR Y}\KET{0}=&e^{-i\pi/4}\KET{0} \quad,\quad
{\BAR Y}{\NOBAR X}{\NOBAR Y}\KET{1}=e^{+i\pi/4}\KET{1} \CR,
{\BAR Y}{\NOBAR X}{\BAR Y}\KET{0}=&e^{+i\pi/4}\KET{1} \quad,\quad
{\BAR Y}{\NOBAR X}{\BAR Y}\KET{1}=-e^{-i\pi/4}\KET{0} \CR,
{\NOBAR Y}{\BAR X}{\BAR Y}\KET{0}=&e^{-i\pi/4}\KET{0} \quad,\quad
{\NOBAR Y}{\BAR X}{\BAR Y}\KET{1}=e^{+i\pi/4}\KET{1} \CR,
}

\EQA{W=&{\NOBAR X}^2{\BAR Y}={\NOBAR Y} {\NOBAR X}^2
=-{\BAR X}^2{\BAR Y}=-{\NOBAR Y} {\BAR X}^2 \CR,
W\KET{0}=&{i\over\sqrt{2}}(\KET{0}+\KET{1}) \quad,\quad
W\KET{1}={i\over\sqrt{2}}(\KET{0}-\KET{1}) \CR,
W^2\KET{0}=&-\KET{0} \quad,\quad
W^2\KET{1}=-\KET{1} \CR.
}

\NEWPAGE
\parskip=0pt plus 0pt
\HEADING{References}\SETREFHEADINGfalse\PRINTREFNOW

\NEWPAGE

\NEXTFIG
\centerline{\vbox to 170true mm{\vfil
\hbox to 150truemm{\hfil\hbox{
\epsffile{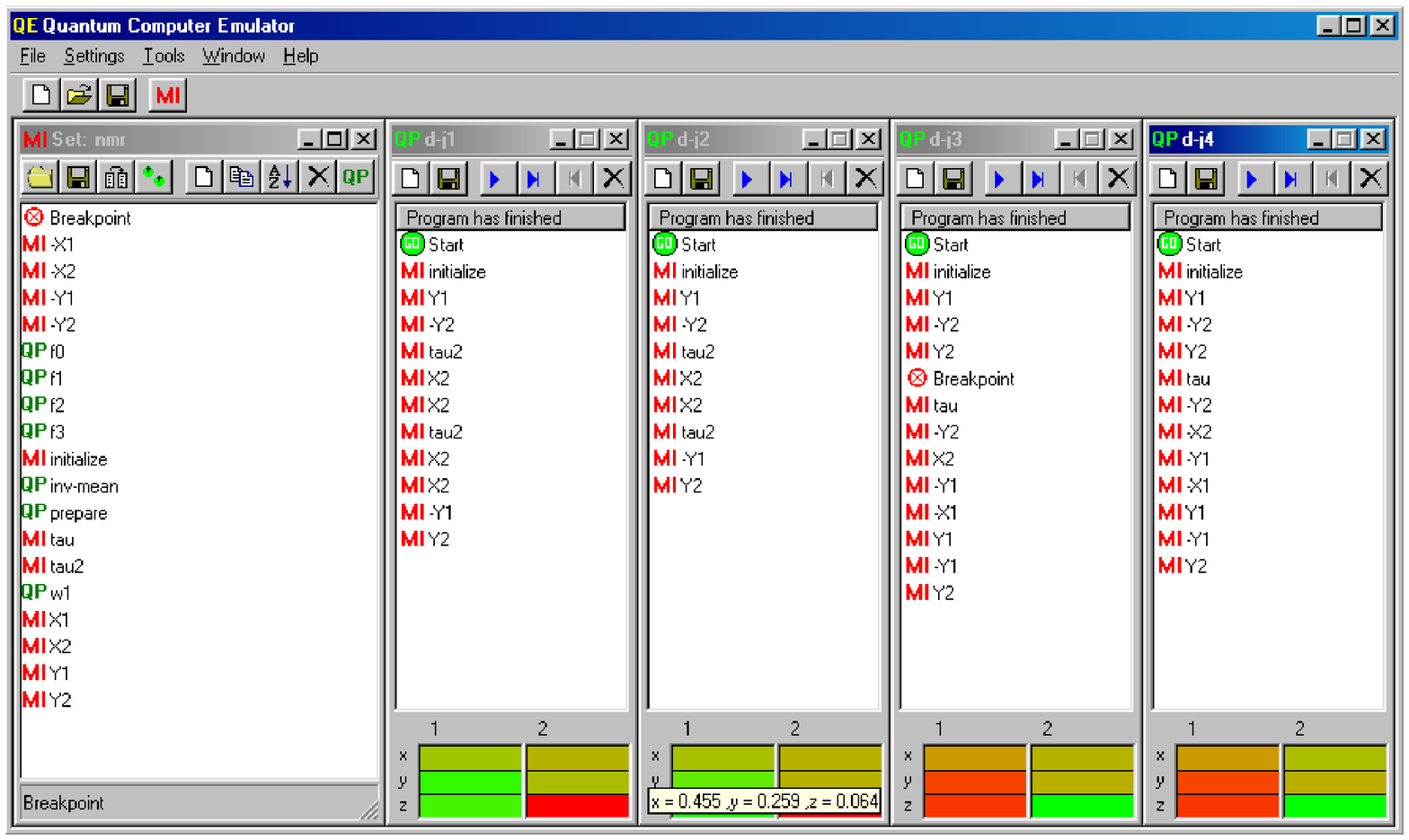}}\hfil}
\vfil 
}}
\medskip\centerline{\vbox{\hsize=150truemm\tenrm\baselineskip=13truept%
\noindent%
Fig.\FIG\ %
Picture of the Quantum Computer Emulator showing a window with a
set of micro instructions implementing an NMR quantum computer and
windows with four Deutsch\-Jozsa programs (d-j1, ..., d-j4),
one for each function ($f_1(x)$, ..., $f_4(x)$) listed in Table 1.
The final state of the QC, i.e. the expectation
value of the qubits (spin operators),
is shown at the bottom of each program
window (green = \KET{0}, red =\KET{1}).
The numerical values appear
if the cursor moves over the qubit area.
}}

\NEWPAGE
\NEXTFIG
\centerline{\vbox to 170true mm{\vfil
\hbox to 150truemm{\hfil\hbox{
\epsffile{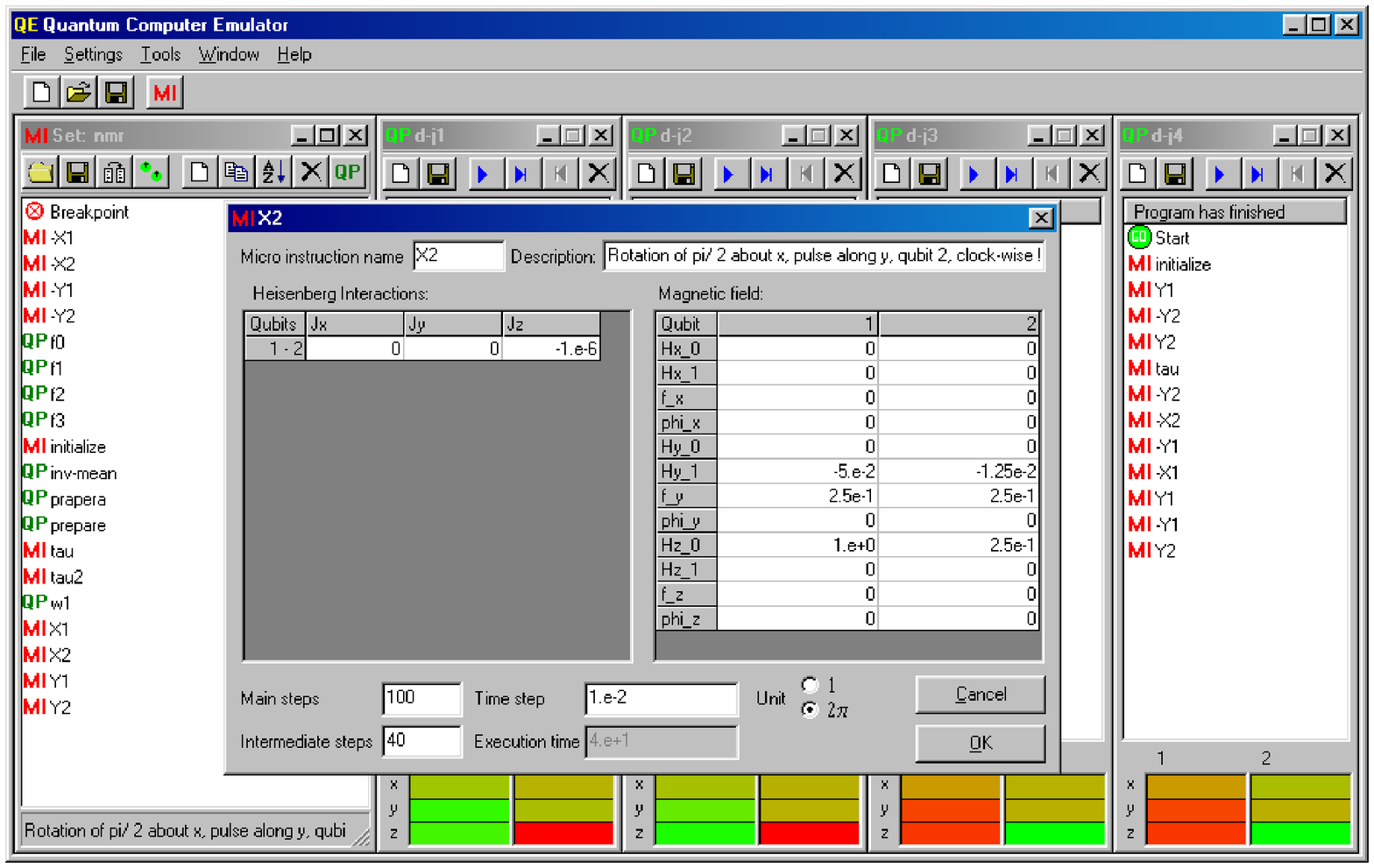}}\hfil}
\vfil 
}}
\medskip\centerline{\vbox{\hsize=150truemm\tenrm\baselineskip=13truept%
\noindent%
Fig.\FIG\ %
Picture of the Quantum Computer Emulator showing a window with a
set of micro instructions implementing an ideal quantum computer and
windows with four Deutsch\-Jozsa programs (d-j1, ..., d-j4),
one for each function ($f_1(x)$, ..., $f_4(x)$) listed in Table 1.
Also shown is a window for editing micro instructions, which
appears by double-clicking on a micro instruction (x2 in this example).
The final state of the quantum computer, i.e. the expectation
value of the qubits (spin operators),
is shown at the bottom of each program
window (green = \KET{0}, red =\KET{1}).
In this ideal case the expectation values are either zero or one.
}}

\NEWPAGE
\NEXTFIG
\centerline{\vbox to 170true mm{\vfil
\hbox to 150truemm{\hfil\hbox{
\epsffile{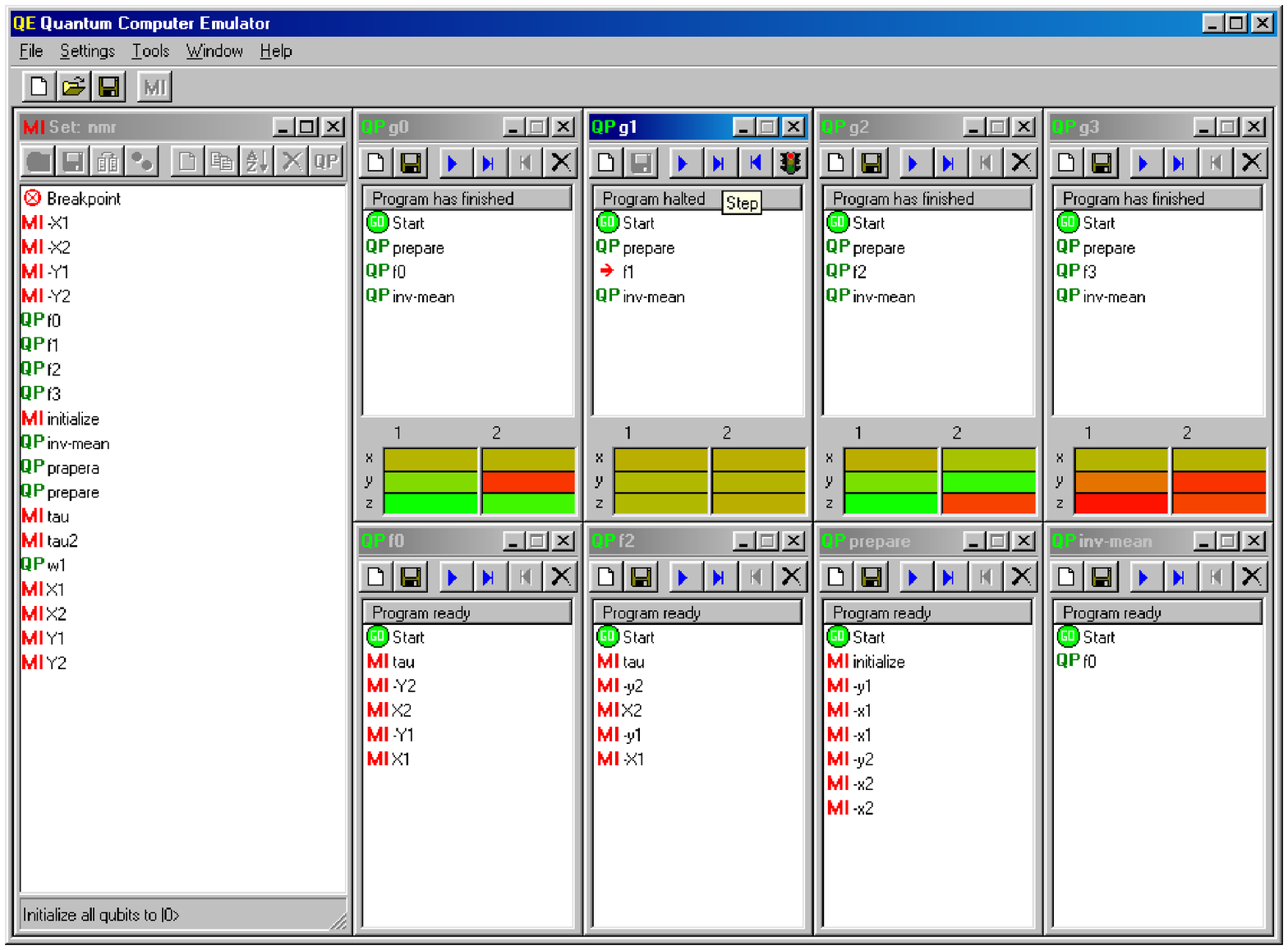}}\hfil}
\vfil 
}}
\medskip\centerline{\vbox{\hsize=150truemm\tenrm\baselineskip=13truept%
\noindent%
Fig.\FIG\ %
Picture of the Quantum Computer Emulator showing a window with a
set of micro instructions for an two\-qubit NMR quantum computer and
windows with quantum programs implementing Grover's
database search for the four different cases
$g_0(x)$ (g0), ..., $g_3(x)$ (g3). This example also shows the use of
quantum programs as micro instructions in other quantum programs.
The final state of the QC, i.e. two qubits shown at the bottom of each program,
gives the location (in binary representation) of the item in the database.
Note that for the case g1 we were using single\-step mode to execute the program
and stopped at f1.
}}

\NEWPAGE
\NEXTFIG
\centerline{\vbox to 170true mm{\vfil
\hbox to 150truemm{\hfil\hbox{
\epsffile{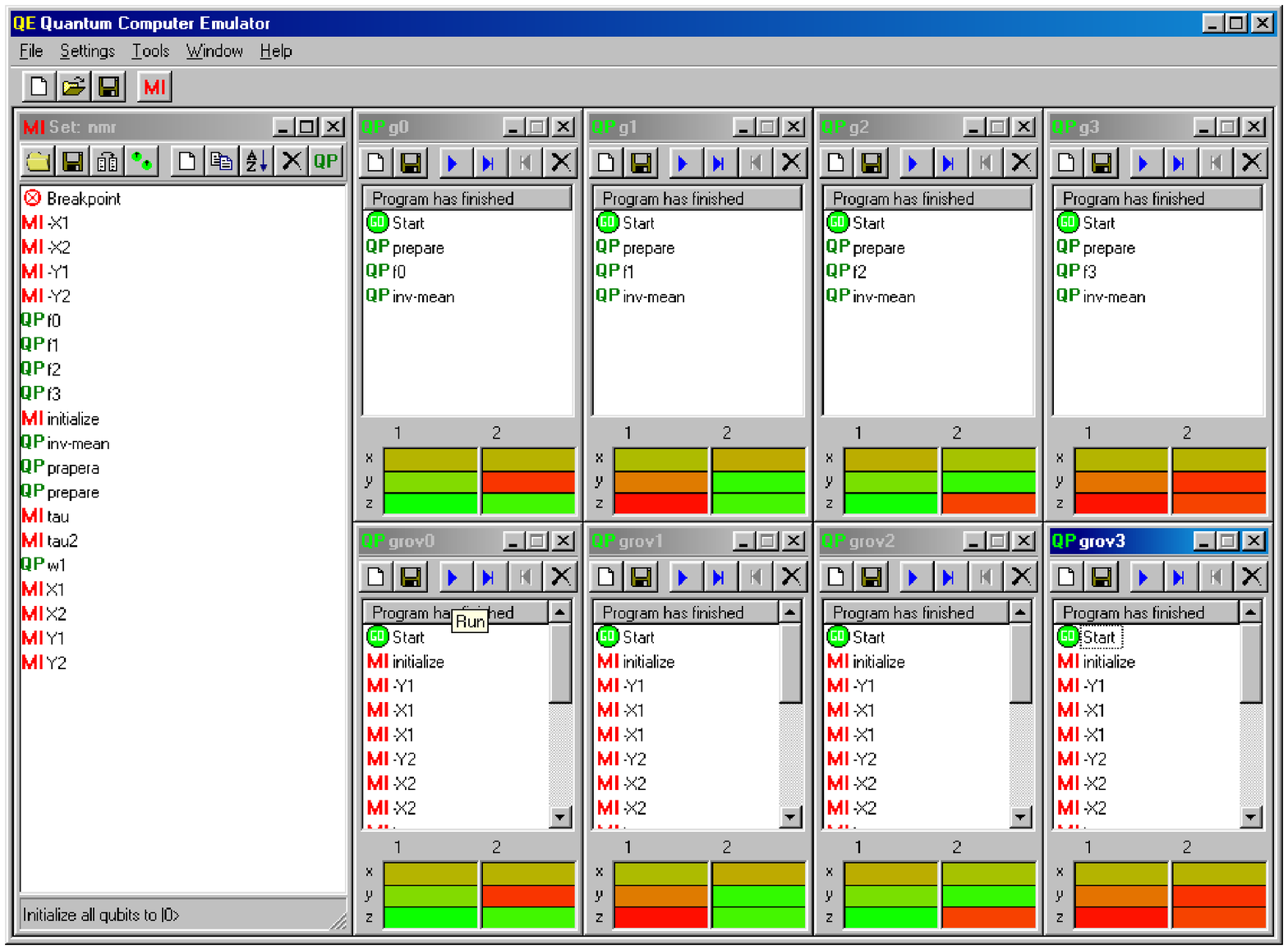}}\hfil} 
\vfil 
}}
\medskip\centerline{\vbox{\hsize=150truemm\tenrm\baselineskip=13truept%
\noindent%
Fig.\FIG\ %
Picture of the Quantum Computer Emulator showing a window with a
set of micro instructions for the two\-qubit NMR quantum computer and
windows with quantum programs implementing Grover's
database search for the four different cases $g_0(x),\ldots\,g_3(x)$, using
the basic MI's (grov0, ..., grov3) and calls to other quantum programs (g0, ..., g3).
The final state of the quantum computer, i.e. two qubits shown at the bottom of each program,
gives the location (in binary representation) of the item in the database.
This final state is no longer a pure basis state but as the weight of
basis state corresponding the location of the item is by far the largest
the correct answer is easy to infer.
}}


\BYE